\documentclass[AMA,STIX1COL]{WileyNJD-v2}
\usepackage{subfigure}
\setcitestyle{square}
\allowdisplaybreaks

\articletype{Research Article}

\begin{document}

\title{Structured Singular Value of a Repeated Complex Full-Block Uncertainty}

\author[1]{Talha Mushtaq}

\author[1]{Diganta Bhattacharjee}

\author[2]{Peter Seiler}

\author[1]{Maziar S. Hemati}

\authormark{Mushtaq \textsc{et al.}}

\address[1]{\orgdiv{Aerospace Engineering and Mechanics}, \orgname{University of Minnesota}, \orgaddress{\city{Minneapolis}, \state{Minnesota}, \country{USA}}}

\address[2]{\orgdiv{Electrical Engineering and Computer Science}, \orgname{University of Michigan}, \orgaddress{\city{Ann Arbor}, \state{Michigan}, \country{USA}}}

\corres{Talha Mushtaq, \email{musht002@umn.edu}}

%\presentaddress{110 Union St SE, 107 Akerman Hall, Minneapolis, MN 55455}
%%%%%%%%%%%%%%%%%%%%%%%%%%%%%%%%%%%
\abstract[Abstract]{
The structured singular value (SSV), or $\mu$, is used to assess the robust stability and performance of an uncertain linear time-invariant system. Existing algorithms compute upper and lower bounds on the SSV for structured uncertainties that contain repeated (real or complex) scalars and/or non-repeated complex full-blocks.
This paper presents algorithms to compute bounds on the SSV for the case of repeated complex full-blocks.
This specific class of uncertainty is relevant for the input-output analysis of many convective systems, such as fluid flows.
Specifically, we present a power iteration to compute the SSV lower bound for the case of repeated complex full-blocks. 
This generalizes existing power iterations for repeated complex scalars and non-repeated complex full-blocks.
The upper bound can be formulated as a semi-definite program (SDP), which we solve using a standard interior-point method to compute optimal scaling matrices associated with the repeated full-blocks.
Our implementation of the method only requires gradient information, which improves the computational efficiency of the method.
Finally, we test our proposed algorithms on an example model of incompressible fluid flow.
The proposed methods provide less conservative bounds as compared to prior results, which ignore the repeated full-block structure.}


\jnlcitation{\cname{%
\author{Talha Mushtaq},
\author{Diganta Bhattacharjee},
\author{Peter Seiler}, and
Maziar S. Hemati} (\cyear{2023}),
\ctitle{Structured Singular Value of a Repeated Complex Full-Block Uncertainty}, \cjournal{Int J Robust Nonlinear Control. } <year> <vol> Page <xxx>-<xxx>}

\maketitle

\footnotetext{\textbf{Abbreviations:} SSV, structured singular value; PCF, plane Couette flow; I/O, input-output}

\section{Introduction} \label{sec:intro}
The structured singular value (SSV), or $\mu$, is a useful metric for assessing the robust stability and performance of an uncertain linear time-invariant system with a structured uncertainty.\cite{doyle1982analysis, packard1993complex,safonov82}
The SSV is  inversely related to the smallest structured uncertainty that destabilizes the uncertain system.
Roughly, the SSV is the ``gain'' of the system with respect to the structured uncertainty and its inverse provides a stability margin.\cite{zhou1996robust,dullerud13}
It is known that exactly computing the SSV is NP hard.\cite{braatz1994computational,demmel92} 
Thus, it is a common practice 
to instead compute upper and lower bounds on the SSV. The upper bound provides a sufficient condition for robust stability and the lower bound for instability, respectively. \cite{zhou1996robust,dullerud13,packard1993complex,young90,young1992}
However, for some specific uncertainty structures, as noted in prior works, \cite{packard1993complex, Troeng2021, colombino2016} the convex upper bound equals the SSV.
Thus, for these cases, the exact SSV can be computed through the convex upper bound.

Much of the previous work has focused on structured uncertainties with a mixture of repeated (real or complex) scalars and/or non-repeated complex full-block uncertainties (see Section \ref{sec:background}).\cite{zhou1996robust,dullerud13,packard1993complex}
For these common uncertainty structures, one can use the methods described in prior works to compute the upper and lower bound.\cite{young1992, Fan1991, young90, packard1988}
The current paper focuses on a new uncertainty structure: repeated complex full-blocks. 
This particular class of uncertainties consists of a single complex full-block repeated multiple times.  
This repeated structure naturally arises in fluid dynamics and other convective systems.  
Recently, SSV has emerged as a means of performing a structured input-output analysis of transitional shear flows to study instability mechanisms.\cite{liu21, liu2022strat, liu2023structured, Bhattacharjee_et_al_2023} 
However, Liu et al. \cite{liu21,liu2022strat,liu2023structured} utilize MATLAB’s Robust Control Toolbox, which does not allow for repeated full-blocks. 
The only cases handled by MATLAB are non-repeated, complex full-blocks and the repeated (real or complex) scalars.\cite{balas2007robust}
Therefore, the numerical results in these works \cite{liu21,liu2022strat,liu2023structured} replace the repeated complex full-block structure with a non-repeated one, which yields conservative SSV bounds.
In addition to conservatism in the bounds, accounting for the repeated uncertainty structure is important for revealing physical instability mechanisms, as will become clear in the results we present later.

In this paper, we present algorithms to compute upper and lower bounds on the SSV for a repeated complex full-block uncertainty (see Sections \ref{sec:upper bound} and \ref{sec:lower bound}).
The upper bound is computed using % a combination of Osborne's iteration and 
an interior point algorithm known as the method of centers.\cite{boyd1993method,boyd1994linear}
Our implementation only uses gradient (and not Hessian) information.  
This improves computational efficiency, which is important for any large dimensioned system, such as the fluid flow example presented in our paper.
The lower bound is computed by generalizing the existing power iteration algorithm described by Packard et al. \cite{packard1993complex, packard1988} 
We demonstrate the proposed algorithms on the plane Couette flow model \cite{liu21} and a simple academic example.
Furthermore, we compare the SSV bounds computed from the proposed algorithms with existing methods that approximate the repeated structure with a non-repeating one.
We show that the proposed algorithms not only reduce the conservatism of the bounds but also highlight the importance of incorporating the correct uncertainty structure for interpreting the underlying physical system/phenomena (see Section \ref{sec:Results}). 

The symbols $\mathbb{R}, \mathbb{C}, \mathbb{R}^n$, $\mathbb{C}^n$ and $\mathbb{C}^{n \times m}$ denote the sets of real numbers, complex numbers, real vectors of dimension $n$, complex vectors of dimension $n$ and complex matrices of dimension $n \times m$, respectively.
The $n\times n$ identity and zero matrices are denoted by $I_n$ and $0_n$, respectively. 
$M^\text{H}$ and $\bar{\sigma}(M)$ are the Hermitian transpose and maximum singular value of a matrix $M \in \mathbb{C}^{n\times m}$.
We use $\|\cdot\|_2$ to denote the 2-norm for vectors and the induced 2-to-2 norm for matrices.
Note that $\|\cdot\|_2 = \bar{\sigma}(\cdot)$ for matrices.
Also, $||\cdot||_F$ denotes the Frobenius norm.
For $M \in \mathbb{C}^{n \times n}$, $\text{Tr}(M)$ and $\rho(M)$ are the trace and spectral radius.
The notations $\otimes$ and $\text{diag}(\cdot)$ denote the Kronecker product and block diagonal matrices, respectively.
The imaginary unit is denoted as $\mathrm{i} =\sqrt{-1}$.
For $c \in \mathbb{C}$, $\text{Re}(c)$, $\text{Im}(c)$ and $\text{conj}(c)$ % $c^*$ 
denote the real and imaginary parts of $c$, and the complex conjugate of $c$, respectively.

\section{Background: Structured Singular Value, $\mu$} \label{sec:background}
We briefly review the structured singular value $\mu$ and its connection to robust stability of dynamical systems. \cite{doyle1982analysis, packard1993complex, safonov1980stability, zhou1996robust}
First consider the case for matrices.
Specifically, let $M\in \mathbb{C}^{n\times m}$ be given along with a set of (possibly structured) complex matrices $\mathbf{\Delta} \subseteq \mathbb{C}^{m\times n}$.  
\begin{definition}
The structured singular value, $\mu_{\mathbf{\Delta}}$, is defined as
\begin{align}
    \mu_{\mathbf{\Delta}}(M) := \frac{1}{\min(\bar{\sigma}(\Delta): \Delta \in \mathbf{\Delta}, \det(I_n - M \Delta) = 0)}.
    \label{eq:ssv}
\end{align}
If there does not exist
$\Delta \in \mathbf{\Delta}$ such that ${\det(I_n - M \Delta) = 0}$, then define $\mu_{\mathbf{\Delta}}(M) = 0$.
\end{definition}

Note that $\mu_{\mathbf{\Delta}}(M)$ depends
on both the matrix $M$ and the set of matrices $\mathbf{\Delta}$.
We will
typically omit the subscript $\mathbf{\Delta}$ for simplicity when the
set of matrices is clear.

The SSV is inversely related to the smallest
$\Delta \in \mathbf{\Delta}$ that causes $I_n - M \Delta$ to be singular.
Singularity means there exists a nonzero vector $y\in \mathbb{C}^n$ such that $y=M\Delta y$.
This is equivalent to the existence of non-zero
vectors $u\in \mathbb{C}^{m}$ and $y \in \mathbb{C}^{n}$ such that $y = Mu$ and
$u = \Delta y$, which provides a feedback interpretation of $\mu_{\mathbf{\Delta}}(M)$ (see Remark 3.4 in Packard and Doyle \cite{packard1993complex}). 
Furthermore, the SSV simplifies in two special cases:\cite{packard1993complex}
\begin{enumerate}[(i)]
    \item $\mu(M) = \bar{\sigma}(M)$ for  full-block uncertainties, $\mathbf{\Delta} = \mathbb{C}^{m\times n}$,
    \item $\mu(M) = \rho(M)$ for repeated scalar uncertainties $\mathbf{\Delta} = \{ \delta I_v : \delta \in \mathbb{C}\}$, where $n,m=v$.
\end{enumerate}
There are many known results for structured uncertainties $\mathbf{\Delta}$ that contain block-diagonal concatenation of any
number of full-blocks and repeated scalars.\cite{doyle1982analysis, packard1993complex, safonov1980stability, zhou1996robust} 
It is worth noting that if $\mathbf{\Delta_1}\subset \mathbf{\Delta_2}$ then
\begin{align}
     \mu_{\mathbf{\Delta_1}}(M) \le \mu_{\mathbf{\Delta_2}}(M).
\end{align}
This follows from the definition of the SSV in \eqref{eq:ssv}. This yields
the following bound for any matrix $M$ and block structure $\mathbf{\Delta}\subseteq \mathbb{C}^{m\times n}$:
\begin{align}
     \mu_{\mathbf{\Delta}}(M) \le \bar{\sigma}(M).
     \label{eq:bound}
 \end{align} 
Next, consider the case for LTI systems.
Specifically, let $M(s)$ be a transfer function matrix of a multiple-input and multiple-output (MIMO) LTI system and $\mathbf{\Delta}$ be a set of structured LTI uncertainties.  
The SSV can be used to assess robustness of a feedback loop involving $M(s)$ and $\Delta(s)$.  
In particular, assume the feedback loop is nominally stable, i.e., stable for $\Delta(s)=0$. 
Define the set of bounded, structured uncertainties as $\mathbb{B}_{\mathbf{\Delta}}:=\{\Delta(s) \in \mathbf{\Delta} \, : \, \| \Delta\|_\infty \le 1\}$.
Then, the feedback loop is stable for all $\Delta \in \mathbb{B}_{\mathbf{\Delta}}$ if and only if $\max_\omega \mu( M(\mathrm{i}\omega) ) <1$, where $\omega$ is the temporal frequency.\cite{doyle1982analysis, packard1993complex, safonov1980stability, zhou1996robust} 
%
%\textcolor{blue}{This is a generalization of the small-gain condition.} 
This is an adaptation of the small-gain condition for the set of structured uncertainties $\mathbb{B}_{\mathbf{\Delta}}$.
%
%\textcolor{blue}{To this end, define the set of full-block uncertainties $\mathbb{C}_{\mathbf{\Delta}}:=\{\Delta(s) \in \mathbb{C}^{m \times n} \, : \, \| \Delta\|_\infty \le 1\}$. 
%%
%The induced $\mathcal{L}_2$ norm of the MIMO LTI system % given by the transfer matrix $M(s)$ 
%is equal to $\max_{\omega} \bar{\sigma}(M(\mathrm{i} \omega))$ and the small-gain condition for all $\Delta \in \mathbb{C}_{\mathbf{\Delta}}$ is $\max_{\omega} \bar{\sigma}(M(\mathrm{i} \omega))<1$. 
%%
%Noting the fact that $\mu$ reduces to $\bar{\sigma}$ for full-block uncertainties $\mathbf{\Delta} = \mathbb{C}^{m\times n}$, the connection becomes clear.}
%%
The SSV computations for LTI systems are often reduced to the SSV computations for a complex matrix $M(\mathrm{i} \omega)$ on a grid of frequencies.

This paper contributes methods that can be used to compute the SSV for repeated full-block uncertainty
\begin{align}
\label{eq:RFB}
  \mathbf{\Delta}_\mathrm{r} := \{ \Delta = I_v \otimes \Delta_1  \, : \, \Delta_1 \in \mathbb{C}^{m_1 \times m_1} \} \subset \mathbb{C}^{m \times m},
\end{align}
where $m=vm_1$.
Thus, $v=2$ represents
the same full-block uncertainty $\Delta_1$ repeated twice:
$I_2 \otimes \Delta_1 = \left[ \begin{smallmatrix} \Delta_1 & 0 \\ 0 & \Delta_1 \end{smallmatrix} \right]$.
The block $\Delta_1$ is restricted to be square, as is common in the SSV literature, to simplify the presentation.  
The extension to non-square blocks can be made with mainly notational changes.
We discuss algorithms in the subsequent sections that compute upper and lower bounds on the $\mu(M)$ for the uncertainty structure in \eqref{eq:RFB}.

%%%%%%%%%%%%%%%%%%%%%%%%%%%%%%%%%%%%%%%%%%%%%%%%%%%%%%%%%%%%%%%%%%%%%%%%%%%%%%%%%%%%%%%%%%%%%%%%%%%%%%%%%%%%%%%%%%%%%%%%%%%%%%%%%%%%%%%%%%%%%%%%%%%%%%%%%%%%%%%%%%%%%%%%%%%%%%%%%%%%%%%%%%%%%%%%%%%%%%%%%%%%%%%%%%%%%%%%%%%%%%%%%%%%%%%%%%%%%%%%%%%%%%%%%%%%%%%%%%%%%%%%%%%

%
\section{Upper Bound of Structured Singular Value} \label{sec:upper bound}
This section describes an algorithm that computes an upper bound on $\mu$ for the uncertainty structure defined in \eqref{eq:RFB}.
We will describe the upper bound algorithm for the matrix case $M \in \mathbb{C}^{m\times m}$.
We start by first noting that for each set of uncertainties $\mathbf{\Delta}$, there is a set of non-singular ``commuting" matrices $\mathbf{D}$ with the property that  $D \Delta = \Delta D$ for any $\Delta \in \mathbf{\Delta}$ and $D \in \mathbf{D}$.
For example, the set of $v$ non-repeated full-blocks,
denoted $\mathbf{\Delta}_\mathrm{nr} \subset \mathbb{C}^{m \times m}$,
and its corresponding commuting matrices are
\begin{align}
   \mathbf{\Delta}_\mathrm{nr} & := \{ \Delta = 
   \text{diag}(\Delta_1, \ldots, \Delta_{v}) : \Delta_i \in \mathbb{C}^{m_i \times m_i}
    \},
    \label{eq:uncert_struct} \\
    \mathbf{D}_\mathrm{nr} & := \{ \text{diag}(d_1 I_{m_1},\ldots, d_v I_{m_v}) : d_i \in \mathbb{R}, d_i \ne 0\}. 
    \label{eq:scaling_struct}
\end{align} 
The commuting matrices are diagonal when the uncertainty set is non-repeated.  For the repeated full-block structure in \eqref{eq:RFB}, the commuting matrices have the following structure:
\begin{align}
      \mathbf{D}_\mathrm{r} := \{ S \otimes I_{m_1}  \, : \,  S \in \mathbb{C}^{v \times v}, \, \det(S)\ne 0 \}.
      \label{eq:scaling_struct_2}
\end{align} 
These commuting matrices are important because 
$\det(I-M \Delta)=\det(I- D M D^{-1} \Delta)$.  Thus, $\mu_{\mathbf{\Delta}}(M) =\mu_{\mathbf{\Delta}}(D M D^{-1})$.
We can use this to strengthen the upper bound in \eqref{eq:bound}:
\begin{align}
    \mu_{\mathbf{\Delta}}(M) \le \min_{D \in \mathbf{D}}  \bar{\sigma}( D M D^{-1}  ).
    \label{eq:max_sing_val}
\end{align} 
This is known as the $D$-scale upper bound.
By setting $X = D^\text{H} D$, the optimization on the right hand side of \eqref{eq:max_sing_val} can be converted into a semi-definite program  (technically a generalized eigenvalue problem) as follows:\cite{packard1993complex, zhou1996robust} 
\begin{align}
    \begin{split}
    & \min_{X = X^\text{H}\in \mathbb{C}^{m \times m}, \ \zeta \in \mathbb{R}} \zeta \\
    & \text{subject to:}\,\,\,  M^{\text{H}}X M < \zeta X, \; X > 0.
    \end{split}
    \label{eq:gevp}
\end{align} 
Then, the upper bound is computed as $ \alpha = (\zeta)^{1/2}$ and the corresponding scale as $D = X^{1/2}$.
Therefore, there is an implicit constraint that $\zeta \geq 0$, which arises naturally during the derivation of constraints in \eqref{eq:gevp} (see Packard and Doyle \cite{packard1993complex} for details). 
The optimization problem \eqref{eq:gevp} can be solved using several existing methods such as method of centers, interior-point methods for linear fractional programming, and primal-dual methods.\cite{boyd1994linear,nesterov1995interior,mehrotra1992implementation}
These methods are efficient for moderate-sized problems but can be computationally costly for larger dimensioned problems. 
Specifically, primal-dual methods tend to be  slower because they require second-order schemes to solve \eqref{eq:gevp}.
Certainly, there are faster algorithms %\footnote{See Appendix \ref{sec:appendix} for a fast algorithm for $D \in \mathbf{D}_\mathrm{r}$} 
that utilize a weaker bound, i.e., $\overline{\sigma}(DMD^{-1}) \leq \|DMD^{-1}\|_F$, which is often sufficient for most large-dimensioned problems.
% given any matrix $M$:
%
In this case, an upper bound for a given matrix $M$ becomes
\begin{align}
\label{eq:osbopt}
    \mu_{\mathbf{\Delta}}(M) \le \min_{D \in \mathbf{D}}  \| D M D^{-1}\|_F.
\end{align} 
See Appendix \ref{sec:appendix} for a fast algorithm for computing an upper bound of the form \eqref{eq:osbopt} for $D \in \mathbf{D}_\mathrm{r}$, i.e., the repeated full-blocks case.
However, using a weaker bound yields conservative estimates of the % SSV 
upper bounds, which can result in large gaps between upper and lower bounds.
The goal of this paper is to present an efficient algorithm that would yield the least conservative upper bounds for $\Delta \in \mathbf{\Delta}_\mathrm{r}$.
Thus, we will implement the method of centers for upper bound calculations, since it is a relatively fast first-order method with good convergence properties.\cite{boyd1993method}
First, we will briefly summarize an existing upper bound method for the uncertainty structure $\mathbf{\Delta}_\mathrm{nr}$, which we will use later to compare with the upper bounds obtained for $\mathbf{\Delta}_\mathrm{r}$. 
%%%%%%%%%%%%%%%%%%%%%%%%%%%%%%%%%%%%%%%%%%%%%%%%%%%%%%%%%%%%%%%%%%%%%%%%%%%%%%%%%%%%%%%%
%
\subsection{Standard Osborne's Method: Non-Repeated Complex full-blocks} \label{sec:standard Osborne}

Osborne's iteration can be used to efficiently solve the optimization problem in the right-hand side of \eqref{eq:osbopt} for specific block structures.\cite{osborne_iter}
For example, a version of Osborne's iteration can be applied to  the structure $\mathbf{\Delta}_\mathrm{nr}$ with scalings $\mathbf{D}_\mathrm{nr}$.
Let $D_{i} \in \mathbf{D}_\mathrm{nr}$ denote a scaling with $d_j=1$ for all $j\ne i$. For example, if $i = 1$ then $d_{1}$ is a variable and $d_j=1$ for $j \neq 1$. In addition, partition $M$
into $m_i\times m_j$ sub-blocks, denoted $\hat{M}_{ij}$, consistent with the block dimensions in $\mathbf{\Delta}_\mathrm{nr}$.
Then, the Frobenius norm can be written as
\begin{align}
\begin{split}
    \|D_{i} M D_{i}^{-1}\|_F^2 & =
    \sum^{v}_{r = 1, r\ne i}
    \frac{1}{d_{i}^2} \|\hat{M}_{ri}\|^2_F
    + d_{i}^2  \|\hat{M}_{ir}\|^2_F.
\end{split}
    \label{eq:osborne_fro}
\end{align}
The optimal value $d_i^\star$ that minimizes \eqref{eq:osborne_fro} is given by
\begin{align}
    d_{i}^\star = \left( \frac{\sum_{r = 1, r\ne i}^{v}\|\hat{M}_{ri}\|^2_F}{\sum_{r = 1, r\ne i}^{v}\|\hat{M}_{ir}\|^2_F} \right)^{1/4}.
    \label{eq:osb_min}
\end{align} 
Each $d_i^\star$ is computed from \eqref{eq:osb_min} using $M$ and the corresponding matrix $D^\star$ is determined.
Then, the cost is obtained as $\|M^{[2]}\|_F^2$, where $M^{[2]} = D^\star M D^{\star^{-1}}$. 
The new $D$-scale is then computed from $M^{[2]}$ and the corresponding new cost is determined.
Thus, the iteration proceeds by updating the matrix as $M^{[k]} = (D^\star)^{[k]} M (D^{\star^{-1}})^{[k]}$ and computing the corresponding $(D^\star)^{[k]}$ until $\|M^{[k]}\|_F^2$ has converged. 
The final $D$-scale is denoted by $D_\mathrm{nr}^\star$ after all the iterations.
Osborne showed that the iterative method always converges to the optimal solution of $\min_{D \in \mathbf{D}}  \| D M D^{-1}\|_F$ for the uncertainty $\mathbf{\Delta}_\mathrm{nr}$ with $m_i=1$.\cite{osborne_iter}
%
%%%%%%%%%%%%%%%%%%%%%%%%%%%%%%%%%%%%%%%%%%%%%%%%%%%%%%%%%%%%%%%%%%%%%%%%%%%%%%%%%%%
\subsection{Method of Centers: Repeated Complex Full-Blocks} \label{sec:moc}
In this section, we discuss the method of centers approach for solving the generalized eigenvalue problem \eqref{eq:gevp} %.
for the case when $\Delta \in \mathbf{\Delta}_\mathrm{r}$ and, consequently, $D \in \mathbf{D}_\mathrm{r}$. 
In this case, we have $X = (S \otimes I_{m_1})^\text{H} (S \otimes I_{m_1}) = S^\text{H}S \otimes I_{m_1} = R \otimes I_{m_1}$, where $R := S^\text{H}S$. 
%
%The constraint $X>0$ in \eqref{eq:gevp} can then be expressed as $R>0$.
%
Therefore, the generalized eigenvalue problem (GEVP) in \eqref{eq:gevp} becomes % the following:
\begin{align}
    \begin{split}
    & \min_{R=R^\text{H} \in \mathbb{C}^{v \times v}, \ \zeta \in \mathbb{R}} \zeta \\
    & \text{subject to:}\,\,\,  M^{\text{H}} (R \otimes I_{m_1}) M < \zeta (R \otimes I_{m_1}),\,\, R > 0.
    %& \qquad \qquad \quad R \geq r_{\min} I_v,\ R \leq r_{\mathrm{cond}} I_v
    \end{split}
    \label{eq:gevp_1}
\end{align}
%where $X=R \otimes I_{m_1}$.
%
%Since a feasible $R$ for \eqref{eq:gevp_1} is scale-invariant (i.e., for a feasible $R$, any $cR$ with $c > 0$ is also feasible), we will replace the $R > 0$ constraint in \eqref{eq:gevp_1} with $I_v \leq R \leq r_{\mathrm{cond}} I_v$ to prevent solutions from becoming ill-conditioned, where $r_{\mathrm{cond}}$ is the (specified) condition number of $R$.
%
Since a feasible $R$ for \eqref{eq:gevp_1} is scale-invariant (i.e., for a feasible $R$, any $cR$ with $c > 0$ is also feasible), we will replace the $R > 0$ constraint in \eqref{eq:gevp_1} with $\frac{1}{\gamma} I_v \leq R \leq \gamma I_v$ to prevent solutions from becoming ill-conditioned, where $\gamma>0$ and $\gamma^2$ is the (specified) condition number of $R$.
Therefore, we numerically implement the following GEVP:
%\begin{align}
%    \begin{split}
%    & \min_{R=R^\text{H} \in \mathbb{C}^{v \times v}, \ \zeta \in \mathbb{R}} \zeta \\
%    & \text{subject to:}\,\,\,  M^{\text{H}} (R \otimes I_{m_1}) M < \zeta (R \otimes I_{m_1}),\,\, \\
%    & \qquad \qquad \quad I_v \leq R \leq r_{\mathrm{cond}} I_v.
%    \end{split}
%    \label{eq:gevp_2}
%\end{align}
%
\begin{align}
    \begin{split}
    & \min_{R=R^\text{H} \in \mathbb{C}^{v \times v}, \ \zeta \in \mathbb{R}} \zeta \\
    & \text{subject to:}\,\,\,  M^{\text{H}} (R \otimes I_{m_1}) M < \zeta (R \otimes I_{m_1}),\,\, \\
    & \qquad \qquad \quad \frac{1}{\gamma}  I_v \leq R \leq \gamma I_v.
    \end{split}
    \label{eq:gevp_2}
\end{align}
The method of centers is an interior-point algorithm that solves for the analytic center of linear matrix inequality (LMI) constraints, given an initial feasible solution.\cite{boyd1993method, boyd1994linear}
Specifically in \eqref{eq:gevp_2}, we are minimizing the largest generalized eigenvalue $\zeta$ of the matrix pair $\left(M^{\text{H}} (R \otimes I_{m_1}) M, (R \otimes I_{m_1})\right)$.
The algorithm utilizes a gradient descent approach, which involves computing the stepping direction towards an optimal $R$ and the smallest $\zeta \geq 0$ satisfying the LMI constraints.
To this end, the directional derivative is computed using a barrier-function for symmetric positive semi-definite matrices, i.e, $J(R) = -\text{log~det}(R)$. 

Next, we will compute the derivative of $J(R)$. 
Let $r_{ij}\in\mathbb{C}$ denote the $(i,j)$ entry of $R$.
Since $R$ is Hermitian, the diagonal entries are real, i.e., $r_{ii}\in \mathbb{R}$.
Note that the derivative of the barrier function is calculated with respect to the real and imaginary parts of each $(i,j)$ element of $R$.
Therefore, each matrix variable in \eqref{eq:gevp_2} is decomposed as a summation in terms of its basis as
$R = \sum_{i,j} r_{ij} R_{ij}$, 
where $R_{ij}$ is the standard basis for $\mathbb{R}^{v \times v}$.
Then, the barrier function and its derivative with respect to $r_{ij}$ are given by
\begin{align}
    J(R) =& -\text{log~det}(L_1) -\text{log~det}(L_2) -\text{log~det}(L_3),  \\
    \frac{\partial J(R)}{\partial r_{ij}} =& -\zeta \text{Tr}\left((R_{ij} \otimes I_{m_1})^{\text{T}} L_1^{-1} \right) \nonumber \\
    & + \text{Tr} \left((R_{ij} \otimes I_{m_1})^{\text{T}} M L_1^{-1} M^{\text{H}} \right) \label{eq:barrier_deriv} \\ 
    & + \text{Tr}\left(R^{\text{T}}_{ij} L_2^{-1} \right) -\text{Tr}\left(R^{\text{T}}_{ij} L_3^{-1} \right), \nonumber
\end{align}
%
%where $ L_1 = \zeta (R \otimes I_{m_1}) - M^\text{H} (R \otimes I_{m_1}) M$, $L_2 = r_{\mathrm{cond}} I_v - R$ and $L_3 = R - I_v$.
%
where $ L_1 = \zeta (R \otimes I_{m_1}) - M^\text{H} (R \otimes I_{m_1}) M$, $L_2 = \gamma I_v - R$ and $L_3 = R - \frac{1}{\gamma}  I_v$.
To further simplify the expression in \eqref{eq:barrier_deriv}, it will be useful to block partition a given matrix $H \in \mathbb{C}^{m\times m}$, where $(H)_{ij} \in \mathbb{C}^{m_1 \times m_1}$ denotes the $(i,j)$ block for all $i, j = 1, \ldots, v$.\footnote{For $m = v$, $(H)_{ij} \in \mathbb{C}$ is the $(i,j)$ scalar element of $H$}
Thus, $\text{Tr}((R_{11} \otimes I_{m_1})^{\text{T}} L_1 ^{-1}) = \text{Tr}((L_{1}^{-1})_{11})$, which can be
generalized to any $(i,j)$, i.e., $\text{Tr}((R_{ij} \otimes I_{m_1})^{\text{T}} L_1 ^{-1}) = \text{Tr}((L_{1} ^{-1})_{ij})$.
The other terms in \eqref{eq:barrier_deriv} can be simplified in a similar manner and we eventually obtain the following expression:
\begin{equation*}
\begin{aligned}
    \frac{\partial J(R)}{\partial r_{ij}} =& -\zeta \text{Tr}\left((L^{-1}_{1})_{ij} \right) + \text{Tr} \left((M L_1^{-1} M^{\text{H}})_{ij} \right) \\ 
    & + (L^{-1}_{2})_{ij}  - (L^{-1}_{3})_{ij}.
\end{aligned}
\end{equation*}
Thus, the derivative $\Phi_R := \partial J / \partial R$ can be expressed as
\begin{equation*}
        \Phi_R = -\zeta \Gamma \left(L_{1}^{-1} \right) + \Gamma \left(M L_1^{-1} M^{\text{H}} \right) + L_{2}^{-1} - L_{3}^{-1},
\end{equation*}
where $\Gamma: \mathbb{C}^{m\times m} \rightarrow \mathbb{C}^{v \times v}$ is a block-trace operator such that the $(i,j)$ entry of  $\Gamma(H)$ is equal to $\text{Tr}\left( (H)_{ij} \right)$.
An overall summary of the proposed method for upper bound calculation using the method of centers is provided in Algorithm \ref{alg:moc}. 

It is possible to compute the convergence properties of the algorithm using a second-order primal dual method, which utilizes the Hessian of the LMIs.
However, second-order methods are computationally expensive, especially when the system has a large dimension.
For practical purposes, it is computationally efficient to first calculate the lower bounds $\beta$ using the power-iteration (see Section \ref{sec:lower bound} for details) and then compute the upper bounds $\alpha$.
Despite the inherent convergence issues of the power-iteration,\cite{young90, packard1988} it is always possible to obtain a perturbation, which would result in a valid lower bound of SSV.
Then, the gap between the upper and lower bound can be studied to assess the viability of the solution.
Therefore, we terminate our algorithm when the upper bounds $\alpha$ are within a certain desired ratio of the lower bounds $\beta$, i.e., $\frac{\alpha}{\beta} \leq p$, where $p > 1$ is the chosen bound of the ratio.
%
% \textcolor{blue}{We choose $p = 1.05$ as the desired ratio for our algorithm to get the bounds within 5\% of one another.}
For example, we can choose $p = 1.05$ as the desired ratio for our algorithm to get the bounds within 5\% of one another.
It is important to note that for the cases where the upper bounds fail to satisfy $p$, we take the next best upper bound that will result in a ratio closest to $p$.
Certainly, if the gap is too large, e.g., $2p < \frac{\alpha}{\beta}$, then either the lower bound has not converged or possibly the upper bound is not exact.
Additionally, a simple
initial estimate of $R$ for Algorithm \ref{alg:moc} is $ R = \text{diag}((d_1^{\star})^2, \ldots, (d_v^{\star})^2)$, where $d_i^{\star}$ is computed from the Osborne's iteration, which we will use in Section \ref{sec:Results} for the results.
\begin{algorithm}[!hbt]
\caption{Upper Bound: Method of Centers} \label{alg:moc}
\begin{algorithmic}[1]
  \State (Initialization) 
  Choose any feasible $\theta \ll 1$, $\epsilon \ll 1$ and $r_{\mathrm{cond}} > 0$. Set $R = \text{diag}((d_1^{\star})^2, \ldots, (d_v^{\star})^2)$, $\alpha = \overline{\sigma}((R \otimes I_{m_1})^{1/2} M (R \otimes I_{m_1})^{-1/2})$ and $\lambda = \alpha + \epsilon$. Choose a suitable $p>1$ and maximum number of iterations $k_m$.
  %%%%%%%%%%%%%%%%%%%%%%
%  \While{\textcolor{blue}{$\frac{\alpha}{\beta} \geq 1.05$ \& $k \leq 500$}} 
%%%%%%%%%%%%%%%%%%%%
\While{$\frac{\alpha}{\beta} \geq p$ \& $k \leq k_m$}
  %%%%%%%%%%%%%%%%%%%%%%
  \State Set $\lambda = (1- \theta) \alpha + \theta \lambda$ and $l = 1$.
  \While{$l \leq 2$}
  %--------------------------------------------------------------
  \State \begin{varwidth}[t]{0.85\linewidth} $L_1 = \lambda^2 (R \otimes I_{m_1}) - M^\text{H} (R \otimes I_{m_1}) M$, $L_2 = \gamma I_v - R$ and $L_3 = R - \frac{1}{\gamma} I_v$. % $L_2 = r_{\mathrm{cond}} I_v - R$, $L_3 = R - I_v$.
   \end{varwidth}
 %--------------------------------------------------------------
  \State \begin{varwidth}[t]{0.85\linewidth} %Let $M_1 = M L_1^{-1} M^\text{H}$. 
  $\Phi_R = \Gamma(M L_1^{-1} M^\text{H}) - \lambda^2 \Gamma(L_1^{-1}) + L_2^{-1} - L_3^{-1}$.
  \end{varwidth} 
  %--------------------------------------------------------------
  \State \begin{varwidth}[t]{0.85\linewidth} Obtain the step-size $\delta$ through a line search.
  \end{varwidth}
  %--------------------------------------------------------------
  \State Set $R = R - \delta \Phi_R$, $l=l+1$. 
  \EndWhile
  %--------------------------------------------------------------
  \State Set $D = (R \otimes I_{m_1})^{1/2}$, $k=k+1$.
  \State Then, $\alpha = \sqrt{\lambda_{\max}(D^{-\text{H}} M^\text{H} (R \otimes I_{m_1}) M D^{-1})}$.
  %--------------------------------------------------------------
\EndWhile
\State The upper bound: $\alpha$
\end{algorithmic}
\end{algorithm}
It should be noted that a variant of Algorithm \ref{alg:moc} can be conceived for $\Delta \in \mathbf{\Delta}_\mathrm{nr}$ by restricting $R$ to be diagonal with real entries.
\section{Lower Bound of Structured Singular Value} \label{sec:lower bound}
In this section, we give details on the computation of SSV lower bound for $\Delta \in \mathbf{\Delta}_\mathrm{r}$ using the generalized power iteration algorithm.
The algorithm follows the same steps as the standard power iteration commonly used for complex uncertainties given in Packard and Doyle \cite{packard1993complex} but with slightly modified equations.
We will show that the generalized version reduces to the standard algorithm for the commonly used complex uncertainties as a special case.
Thus, the standard power iteration for the repeated scalars and full-block uncertainties is described first so the extension to the generalized version will be clear.
\subsection{Standard Power Iteration: Repeated Scalars and Full Blocks}
This section briefly summarizes the SSV power iterations for complex uncertainties described in Packard and Doyle. \cite{packard1993complex}
We will consider a problem with a given $M\in \mathbb{C}^{m \times m}$ and a block structure with
one repeated scalar and one full-block:
\begin{align*}
  \mathbf{\Delta} := \left\{ 
\Delta = \begin{bmatrix} \delta_1 I_{m_1} & 0 \\ 0 & \Delta_2 \end{bmatrix}
\, : \,  \delta_1 \in \mathbb{C}, \, \Delta_2 \in \mathbb{C}^{m_{2} \times m_{2}} \right\} 
\end{align*}
where, for consistency among the dimensions, we have $m = m_1 + m_2$.
The power iteration will be described for this particular block
structure.
The generalization to other uncertainty block structures
with arbitrary numbers of repeated scalars or full-blocks will be
clear.

Note that any particular $\Delta\in \mathbf{\Delta}$ such that
$\det(I_n-M\Delta)=0$ yields a lower bound
$\mu(M) \ge \frac{1}{\bar\sigma(\Delta)}$.
The exact value of
$\mu(M)$ is computed by finding the ``smallest''
$\Delta \in \mathbf{\Delta}$ such that $\det(I_n - M\Delta)=0$.
The
determinant condition is equivalent to finding
$\Delta \in \mathbf{\Delta}$ and non-zero vectors $y\in \mathbb{C}^{m}$ and
$u\in \mathbb{C}^{m}$ such that $y=Mu$ and $u=\Delta y$.
The power iteration
is an efficient method to find uncertainties
$\Delta \in \mathbf{\Delta}$ that satisfy the determinant condition.
The power iteration does not, in general, find the smallest
uncertainty and hence it only yields a lower bound on
$\mu(M)$. However, these lower bounds are often accurate in
practice. \cite{packard1993complex}
Moreover, the particular uncertainty returned by the power
iteration can be studied further for insight.

To describe the power iteration, consider vectors $a,z,b,w\in \mathbb{C}^{m}$.
Partition these vectors compatibly with the
block structure, e.g.,  $b=\left[ \begin{smallmatrix} b_1 \\ b_2\end{smallmatrix} \right]$ with
$b_1 \in \mathbb{C}^{m_1}$ and $b_2 \in \mathbb{C}^{m_2}$.
The power iteration is
defined based on the following set of equations for some $\beta>0$:
\begin{subequations}
\label{eq:SSValign}
\begin{align}
\label{eq:SSValignA}
& \beta a = Mb \\
\label{eq:SSValignB}
& z_1 = \frac{w_1^\text{H} a_1}{|w_1^\text{H}a_1|} w_1, \,\,
z_2 = \frac{\|w_2\|_2}{\|a_2\|_2} a_2\\
\label{eq:SSValignC}
& \beta w = M^\text{H} z \\
\label{eq:SSValignD}
& b_1 = \frac{a_1^\text{H} w_1}{|a_1^\text{H} w_1|} a_1, \,\,
b_2 = \frac{\|a_2\|_2}{\|w_2\|_2} w_2.
\end{align}
\end{subequations}
These equations arise from the optimality conditions for the SSV and are
related to the concept of principle direction alignment (see details in Packard and Doyle \cite{packard1993complex}).
Here, we will simply show that any solution of
these equations yields a lower bound on $\mu(M)$. First note that \eqref{eq:SSValignD} implies that 
$b_1 = q_1 a_1$ with $q_1:=\frac{a_1^\text{H} w_1}{|a_1^\text{H} w_1|} \in \mathbb{C}$
and $|q_1|=1$.
Equation
\eqref{eq:SSValignD} also gives $\|b_2\|_2 = \|a_2\|_2$. 
Hence, there is a
$Q_2\in \mathbb{C}^{m_2\times m_2}$ with $\bar\sigma(Q_2) = 1$ such
that $b_2 = Q_2 a_2$. 
Finally, define $u:=b$,
$y:=\beta a$ and
$\Delta :=\frac{1}{\beta} \text{diag}(q_1 I_{m_1},\, Q_2)$.
It can be verified from \eqref{eq:SSValignA} that $y=Mu$. Moreover,  $u=\Delta y$ and $\bar\sigma(\Delta)
=\frac{1}{\beta}$ by construction.
Hence, $\Delta\in \mathbf{\Delta}$ satisfies the
determinant condition and yields the lower bound $\mu(M)\geq \frac{1}{\bar\sigma(\Delta)}=\beta$. 

The power iteration attempts to solve \eqref{eq:SSValign} by iterating through the various relations therein. 
The procedure is summarized in Algorithm \ref{alg:piter}. 
The algorithm, as stated, runs for a fixed number of $k_m$ iterations. 
However, more advanced stopping conditions can be used, e.g., terminating when the various vectors have % a 
small updates as measured in the Euclidean norm.
Although $b^{[0]},w^{[0]}$ can be chosen randomly, a more specific choice would be to use the right singular vector associated with % the largest singular value of 
$\bar{\sigma}\left(D_\mathrm{nr}^\star M \left(D_\mathrm{nr}^\star\right)^{-1}\right)$, where $D_\mathrm{nr}^\star$ is obtained using the standard Osborne's iterations \cite{packard1988}.
\begin{algorithm}[!hbt]
\caption{Lower Bound: Power Iteration}
\begin{algorithmic}[1]
%--------------------------------------------------------------------------------------------------------------------------------------
\State (Initialization) Choose the number of iterations $k_m$ and set $k=0$. Select % any 
some unit-norm vectors $b^{[0]},w^{[0]} \in \mathbb{C}^{m}$ and $a^{[0]} = z^{[0]} = 0\in \mathbb{C}^{m}$. 
%--------------------------------------------------------------------------------------------------------------------------------------
\While{$k < k_m$}
%--------------------------------------------------------------------------------------------------------------------------------------
  \State{\eqref{eq:SSValignA}: $\beta:= \|Mb^{[k]}\|_2$ and $a^{[k+1]}:=Mb^{[k]}/\beta$.}
%--------------------------------------------------------------------------------------------------------------------------------------
  \State \eqref{eq:SSValignB}: Use $(a^{[k+1]},w^{[k]})$ to compute $z^{[k+1]}$.
%--------------------------------------------------------------------------------------------------------------------------------------
  \State \eqref{eq:SSValignC}: \begin{varwidth}[t]{0.85\linewidth} $\beta:= \|M^\text{H} z^{[k+1]}\|_2$ and $w^{[k+1]}:=M^\text{H} z^{[k+1]}/\beta$. \end{varwidth}
%--------------------------------------------------------------------------------------------------------------------------------------
  \State \eqref{eq:SSValignD}: Use $(a^{[k+1]},w^{[k+1]})$ to compute $b^{[k+1]}$.
%--------------------------------------------------------------------------------------------------------------------------------------
  \State Set $k = k + 1$.
%--------------------------------------------------------------------------------------------------------------------------------------
\EndWhile
\State Use $a^{[k_m]}$, $b^{[k_m]}$ and $\beta$ to compute $u$, $y$ and $\Delta$.
\end{algorithmic}
\label{alg:piter}
\end{algorithm}
%

% PJS: Need to include a line break after the algorithm. Otherwise
% the next line "This power iteration..." is included in the 
% previous paragraph which looks weird. Instead, we want "This
% power iteration...." to appear after the algorithm and before
% the enumerated list.

This power iteration simplifies in two special cases:
\begin{enumerate}[(i)]
\item $\mathbf{\Delta} = \mathbb{C}^{m\times m}$: As noted above,
  $\mu(M) = \bar{\sigma}(M)$ in this case. The power iteration
  relations in \eqref{eq:SSValign} become
  \begin{align*}
    \beta a = Mb, \,\,\,
    z = \frac{\|w\|_2}{\|a\|_2}a, \,\,\,
    \beta w = M^\text{H} z, \,\,\,
    b = \frac{\|a\|_2}{\|w\|_2}w.
  \end{align*}
  If $b$ and $w$ are initialized to be unit norm, then all vectors are unit norm throughout the iteration. 
  Hence $z=a$ and $b=w$, so the relations further simplify to
  \begin{align*}
    \beta a = M b, \,\,\,
    \beta b = M^\text{H} a. \,\,\,
  \end{align*}
  We can iterate on these equations starting from an initial unit norm vector
  $b$. 
  This corresponds to the standard power iteration for computing
  $\bar\sigma(M)$.

\item $m :=v$ and
  $\mathbf{\Delta} = \{ \delta I_v \, : \, \delta \in \mathbb{C}\}$: As noted above, 
  $\mu(M) = \rho(M)$ in this case. The power iteration
  relations in \eqref{eq:SSValign} simplify to
  \begin{align*}
    \beta a = Mb, \,\,
    z = \frac{w^\text{H} a}{|w^\text{H} a|} w, \,\,
    \beta w = M^\text{H} z, \,\,
    b = \frac{a^\text{H} w}{|a^\text{H} w|} a. \,\,
  \end{align*}
  Iterating these relations yields a power iteration to find the
  eigenvalue corresponding to the spectral radius. The iteration also  yields the
  corresponding right $b$ and left $z^\text{H}$ eigenvectors.
\end{enumerate}

\subsection{Generalized Power Iteration: Repeated Complex Full-Blocks} \label{sec:generalized power-iteration}
This subsection describes a generalization of the SSV power iteration to
handle repeated complex full-blocks. Again, we consider the problem
with $M\in \mathbb{C}^{m \times m}$ and a structured uncertainty
with one $m_1\times m_1$ full-block repeated $v$ times as in \eqref{eq:RFB}.
A lower bound on $\mu(M)$ is obtained by finding
$\Delta \in \mathbf{\Delta}$ and non-zero vectors $y\in \mathbb{C}^{m}$ and
$u\in \mathbb{C}^{m}$ such that $y = Mu$ and $u=\Delta y$.  

It will be useful to define the following reshaping operation
$L_{m_1}:\mathbb{C}^{v m_1} \to \mathbb{C}^{m_1 \times v}$ such that $y=\begin{bmatrix} y_1^\text{H} & \ldots & y_v^\text{H} \end{bmatrix} ^ \text{H} \in  \mathbb{C}^{v m_1}
\mbox{ maps to }
  L_{m_1}(y) = \begin{bmatrix} y_1, \, \ldots, y_v \end{bmatrix}$.
This operation restacks the partitioned vector $y\in \mathbb{C}^{vm_1}$
into a matrix.  
The inverse $L_{m_1}^{-1}$ will convert the matrix
back to a column vector. 
This notation is useful to handle matrix-vector products for $\Delta\in\mathbf{\Delta}$. Specifically, let $\Delta=I_v\otimes \Delta_1$ with $\Delta_1\in \mathbb{C}^{m_1\times m_1}$.
The relation $u=\Delta y$
is equivalent to $L_{m_1}(u) = \Delta_1 L_{m_1}(y)$. 

We need one additional operation to define the generalized power iteration.
Consider vectors
$a, z, b, w\in \mathbb{C}^{v m_1}$.  
Let
$G$ be a matrix of any dimension with the
following SVD:
\begin{align}
G = U \Sigma V^\text{H} 
   = \begin{bmatrix} U_1 & U_2 \end{bmatrix} \begin{bmatrix} \hat{\Sigma} & 0 \\ 0 & 0 \end{bmatrix}
     \begin{bmatrix} V_1 & V_2 \end{bmatrix}^\text{H}.
\end{align}
Define $\mathbf{Q}(G):=U_1 V_1^\text{H}$ and note that $\bar{\sigma}( \mathbf{Q}(G) ) =1$. The
power iteration is defined based on the following set of equations for
some $\beta>0$:
\begin{subequations}
\label{eq:SSValignRFB}
\begin{align}
\label{eq:SSValignRFBA}
& \beta a = M b \\
\label{eq:SSValignRFBB}
& L_{m_1}(z) = \mathbf{Q}\left(L_{m_1}(a)L_{m_1}(w)^\text{H}\right) \,  L_{m_1}(w)\\
\label{eq:SSValignRFBC}
& \beta w = M^\text{H} z \\
\label{eq:SSValignRFBD}
& L_{m_1}(b) = \mathbf{Q}\left(L_{m_1}(w)L_{m_1}(a)^\text{H}\right) \, L_{m_1}(a).
\end{align}
\end{subequations}
Any solution of these equations yields a lower bound on
$\mu(M)$. To show this, define $u := b$, $y:=\beta a$ and
$\Delta :=I_v \otimes
\frac{1}{\beta}\mathbf{Q}\left(L_{m_1}(w)L_{m_1}(a)^\text{H}\right)$. Then
\eqref{eq:SSValignRFBA} and \eqref{eq:SSValignRFBD} are equivalent to
$y = M u$ and $u=\Delta y$. Moreover,
$\bar{\sigma}(\Delta) =\frac{1}{\beta}$ by construction. Hence
$\Delta\in \mathbf{\Delta}$ satisfies the determinant condition and
yields the lower bound $ \mu(M) \geq \beta$.
A power iteration can be used to find a solution by iterating through equations \eqref{eq:SSValignRFBA}--\eqref{eq:SSValignRFBD} as outlined in Algorithm \ref{alg:gen-piter}.
Note that the comments on initialization and stopping criterion for Algorithm \ref{alg:piter} applies for Algorithm \ref{alg:gen-piter} as well.
In cases where the power iteration does not converge, the perturbations $\Delta_1 = I_v \otimes \mathbf{Q}\left(L_{m_1}(a)L_{m_1}(w)^\text{H}\right)$ and $\Delta_2 = I_v \otimes \mathbf{Q}\left(L_{m_1}(w)L_{m_1}(a)^\text{H}\right)$ can be used to obtain a valid lower bound as $\beta = \max\left(\rho(\Delta_1^\text{H} M), \rho(\Delta_2 M)\right)$.
%%%%%%%%%%%%%%%%%%%%%%%%%%%%%%%%%%%%%%%%%%%%%%%%%%%%%%%%%%%%%%%%%%%%%%%%%%%%%%%%%%%%%
\begin{algorithm}[!hbt]
\caption{Lower Bound: Generalized Power Iteration}
\begin{algorithmic}[1]
%--------------------
\State (Initialization) Choose the number of iterations $k_m$ and set $k=0$. Select some unit-norm vectors $b^{[0]},w^{[0]} \in \mathbb{C}^{m}$ and $a^{[0]} = z^{[0]} = 0\in \mathbb{C}^{m}$. 
%---------------------------
\While{$k < k_m$}
%-----------
  \State \eqref{eq:SSValignRFBA}: $\beta:= \| Mb^{[k]} \|_2$ and $a^{[k+1]}:=Mb^{[k]}/\beta$.
%-------------------------------------
  \State \eqref{eq:SSValignRFBB}: \begin{varwidth}[t]{0.8\linewidth} 
  $z_L := \mathbf{Q}\left(L_{m_1}(a^{[k+1]})L_{m_1}(w^{[k]})^\text{H}\right) \\ L_{m_1}(w^{[k]})$ and $z^{[k+1]} = L_{m_{1}}^{-1} (z_L)$
  \end{varwidth}
%------------------------------
  \State \eqref{eq:SSValignRFBC}: \begin{varwidth}[t]{0.85\linewidth} $\beta:= \|M^\text{H} z^{[k+1]}\|_2$ and $w^{[k+1]}:=M^\text{H} z^{[k+1]}/\beta$. \end{varwidth}
%---------------------------------
  \State \eqref{eq:SSValignRFBD}: \begin{varwidth}[t]{0.85\linewidth}
  $b_L := \mathbf{Q}\left(L_{m_1}(w^{[k+1]}) L_{m_1}(a^{[k+1]})^\text{H}\right) \\ L_{m_1}(a^{[k+1]})$ and $b^{[k+1]} = L_{m_{1}}^{-1} (b_L)$.
  \end{varwidth}
%-----------------------------
  \State Set $k = k + 1$.
%----------------------------
\EndWhile
\State Use $a^{[k_m]}$, $b^{[k_m]}$, $w^{[k_m]}$ and $\beta$ to compute $u$, $y$ and $\Delta$.
\end{algorithmic}
\label{alg:gen-piter}
\end{algorithm}
%%%%%%%%%%%%%%%%%%%%%%%%%%%%%%%%%%%%%%%%%%%%%%%%%%%%%%%%%%%%%%%%%%%%%%%%%%%%%%%%%%%%%%%%%
%

% PJS: Again, I added a line break after the algorithm so that 
% "Equations...." appears after the algorithm and right before
% the enumerated list.  (Without the line break this appears
% at the end of the previous paragraph which look weird.).

Equations \eqref{eq:SSValignRFBB} and
\eqref{eq:SSValignRFBC} generalize the cases in the previous subsection:
\begin{enumerate}[(i)]
\item $v=1$:  In this case, the block structure \eqref{eq:RFB}
is just a single full-block uncertainty. The stacking operations
are just $L_{m_1}(z)=z$, $L_{m_1}(a)=a$, $L_{m_1}(w)=w$, and
$L_{m_1}(b)=b$. Thus, an SVD of $L_{m_1}(a)L_{m_1}(w)^\text{H}=aw^\text{H}$
is given by $U_1=\frac{a}{\|a\|_2}$, $V_1=\frac{w}{\|w\|_2}$, and
$\hat{\Sigma} = \|a\|_2 \|w\|_2$. Equation \eqref{eq:SSValignRFBB}
is thus equivalent to $z=\frac{\|w\|_2}{\|a\|_2} a$, which corresponds
to the full-block update in \eqref{eq:SSValignB}.

\item $m_1 =1$: In this case, the block structure
  \eqref{eq:RFB} is a scalar uncertainty repeated $v$ times.  The
  stacking operations are just $L_{m_1}(z)=z^T$, $L_{m_1}(a)=a^T$,
  $L_{m_1}(w)=w^T$, and $L_{m_1}(b)=b^T$. Thus, the stacking operation is
  a transpose (but not conjugation) of the column vector to a row
  vector.  This yields:
  \begin{align}
  L_{m_1}(a)L_{m_1}(w)^\text{H}=a^T (w^T)^\text{H} = w^\text{H} a.       
  \end{align}
  This is a scalar and an SVD of this product is given by
  $U_1=\frac{w^\text{H} a}{|w^\text{H} a|}$, $V_1=1$, and $\hat{\Sigma} = |w^\text{H} a|$. Step
  \eqref{eq:SSValignRFBB} is thus equivalent to
  $z=\frac{|w^\text{H}a|}{|w^\text{H} a|} w$. This corresponds to the repeated scalar
  block update in \eqref{eq:SSValignB}.
\end{enumerate}
%------------------------------------------------------
%%%%%%%%%%%%%%%%%%%%%%%%%%%%%%%%%%%%%%%%%%%%%%%%%%%%%%%%%%%%%%%%%%%%%%%%%%%%%%%%%%%%%%%%%%%%%%%%%%%%%%%%%%%%%%%%%%%%%%%%%%%%%%%%%%%%%%%%%%%%%%%%%%%%%%%%%%%%%%%%%%%%%%%%%%%%%%%%%%%%%%%%%%%%%%%%%%%%%%%%%%%%%%%%%%%%%%%%%%%%%%%%%%%%%%%%%%%%%%%%%%%%%%%%%%%%%%%%%%%%%%%%%%%
%
\section{Results}\label{sec:Results}
We consider a fluid-flow problem wherein the uncertainty has a repeated full-block structure as in \eqref{eq:RFB}. 
The SSV bounds are computed for the true uncertainty structure (i.e., $\Delta \in \mathbf{\Delta}_\mathrm{r}$) using the proposed methods.
We compare those bounds with the ones obtained by treating the uncertainty to be non-repeating (i.e., $\Delta \in \mathbf{\Delta}_\mathrm{nr}$), which is an approximation of the true uncertainty. 
The motivation behind this comparison is to highlight the differences that arise due to this approximation, and how those differences can alter the subsequent interpretation of the physical system/phenomena.
The algorithms used for different cases are summarized in Table \ref{table:different algorithms}. 
Furthermore, we showcase the gap between the upper and lower bounds for the two sets of results. 
In addition, all of the above mentioned aspects have been repeated for a simple academic example.
%%%%%%%%%%%%%%%%%%%%%%%%%

\begin{table}[htb]
\centering
\caption{Different algorithms used for the results \label{table:different algorithms}}%
% \begin{tabular*}{\textheight}{@{\extracolsep\fill}lcc@{\extracolsep\fill}}
\begin{tabular}{ccc}
\toprule
\textbf{Uncertainty Structure} & $\Delta \in \mathbf{\Delta}_\mathrm{nr}$  & $\Delta \in \mathbf{\Delta}_\mathrm{r}$ \\
\midrule
\qquad \textbf{Upper Bound Algorithm} & Osborne's iteration (Section \ref{sec:standard Osborne}) & Method of centers (Algorithm \ref{alg:moc}) \\
\midrule
\qquad \textbf{Lower Bound Algorithm} & Power iteration (Algorithm \ref{alg:piter})  & Generalized power iteration (Algorithm \ref{alg:gen-piter}) \enspace \enspace
 \\
\bottomrule
\end{tabular}
\end{table}

%%%%%%%%%%%%%%%%%%%%%%%%%
\subsection{Example Model-1: Incompressible Plane Couette Flow} \label{Sec:Example Model-1}

We will demonstrate our proposed algorithms on the same spatially discretized incompressible plane Couette flow (PCF) model initially used to investigate SSV---with non-repeated full-blocks---in Liu and Gayme. \cite{liu21}
PCF is a simple shear-driven flow between two parallel plates, wherein the lower plate is held stationary and the upper plate moves with a fixed speed~$U_\infty$.
The PCF example is chosen as a demonstration in this study, but the proposed methods are equally applicable to other systems where repeated full-block uncertainties arise.\cite{liu21,liu2022strat,McKeonJFM2010,chavarin2020resolvent}

The input-output (I/O) map of the forced perturbation dynamics about a steady baseflow is a frequency response matrix defined as 
\begin{align}
     M= C_{\nabla}(\kappa_x,\kappa_z)(\mathrm{i}\omega I_{2s} - A(\kappa_x,\kappa_z))^{-1}B(\kappa_x,\kappa_z),
     \label{eq:transfunc}
\end{align} 
where $\omega$ is the temporal frequency, $\kappa_x$ and $\kappa_z$ are the wavenumbers from discretization in $x$ and $z$ directions using Fourier modes, and $A(\kappa_x, \kappa_z, Re) \in \mathbb{C}^{2s \times 2s}$, $B(\kappa_x, \kappa_z) \in \mathbb{C}^{2s \times m}$ and $C_\nabla (\kappa_x, \kappa_z) \in \mathbb{C}^{n \times 2s}$ are the system operators, respectively.
Additionally, $A$ is a function of the Reynolds number $Re = U_\infty h/\nu$, where $h$ is the distance between the two plates and $\nu$ is the kinematic viscosity of the fluid.
Then, for $M \in \mathbb{C}^{n \times m}$, we simply have the relation $\eta = Mf$ between the system inputs $(f(y,t) \in \mathbb{C}^{m})$ and outputs $(\eta(y,t) \in \mathbb{C}^{n})$ defined as
\begin{align*}
\begin{split}
f = \begin{bmatrix} f_x(y,t) \\ f_y(y,t) \\ f_z(y,t)
\end{bmatrix}, \
\eta(y,t) = \begin{bmatrix} \nabla u_x(y,t) \\ \nabla u_y(y,t) \\ \nabla u_z(y,t)
\end{bmatrix},
\end{split}
\end{align*}
where $\nabla$ is the discrete gradient operator, $u_x$, $u_y$ and $u_z$ represent flow perturbation velocities and $f_x$, $f_y$ and $f_z$ represent input forcing, in $x$, $y$ and $z$ directions, respectively. 
The forcing signal $f$ is a pseudo-linear approximation of the quadratic convective nonlinear term in the incompressible PCF model.
%
%Specifically, $f = \Delta \eta$, which is an approximation of the $\begin{bmatrix} u_x & u_y & u_z \end{bmatrix}^\text{T} \cdot \eta$ term and thus $\Delta_1 = \begin{bmatrix} \hat{u}_x & \hat{u}_y & \hat{u}_z \end{bmatrix}$, where $(\hat{\cdot})$ denotes a constant unknown matrix approximation of each velocity term.
%
This is given by $f = \Delta \eta$ with $\Delta = I_3 \otimes \Delta_1$, where $\Delta_1$ is considered an unknown matrix approximation of the velocity vectors (see Liu and Gayme \cite{liu21} for more details).
Therefore, the uncertainty for this system is of the form shown in \eqref{eq:RFB} with a rectangular block $\Delta_1 \in \mathbb{C}^{m_1 \times n_1}$ repeated three times $(v = 3)$.
%
% Structured singular values 
Thus, the SSV bounds for the PCF model % allow us to determine 
indicate the sensitivity of flow at each $\kappa_x$ and $\kappa_z$ to this forcing, which is an indication of flow's potential for transition to turbulence.\cite{liu20}
Large bound values indicate that the system in \eqref{eq:transfunc} has a higher tendency to transition, and vice versa, which is a consequence of a variation of the small-gain condition for structured uncertainties (see Section \ref{sec:background}).
% a consequence of the small-gain theorem.\cite{zhou1996robust} }
%
For additional details on the model formulation and discretization, we refer the reader to prior works.\cite{liu21, jovanovic2005componentwise}

%%%%%%%%%%%%%%%%%%%%%%%%%%%%%%%%%%%%%%%%%%%%%%%%%%%%%%%%%%%%%%%%%%%%%%%%%%%%%%%%%%%%%%%%%
\begin{figure*}[!hbt]
    \begin{center}
    \subfigure[$\log_{10} (\alpha_{\max})$ for $\Delta \in \mathbf{\Delta}_\mathrm{nr}$: Osborne's iteration (Upper Bounds)]{\includegraphics[width = 0.485\textwidth]{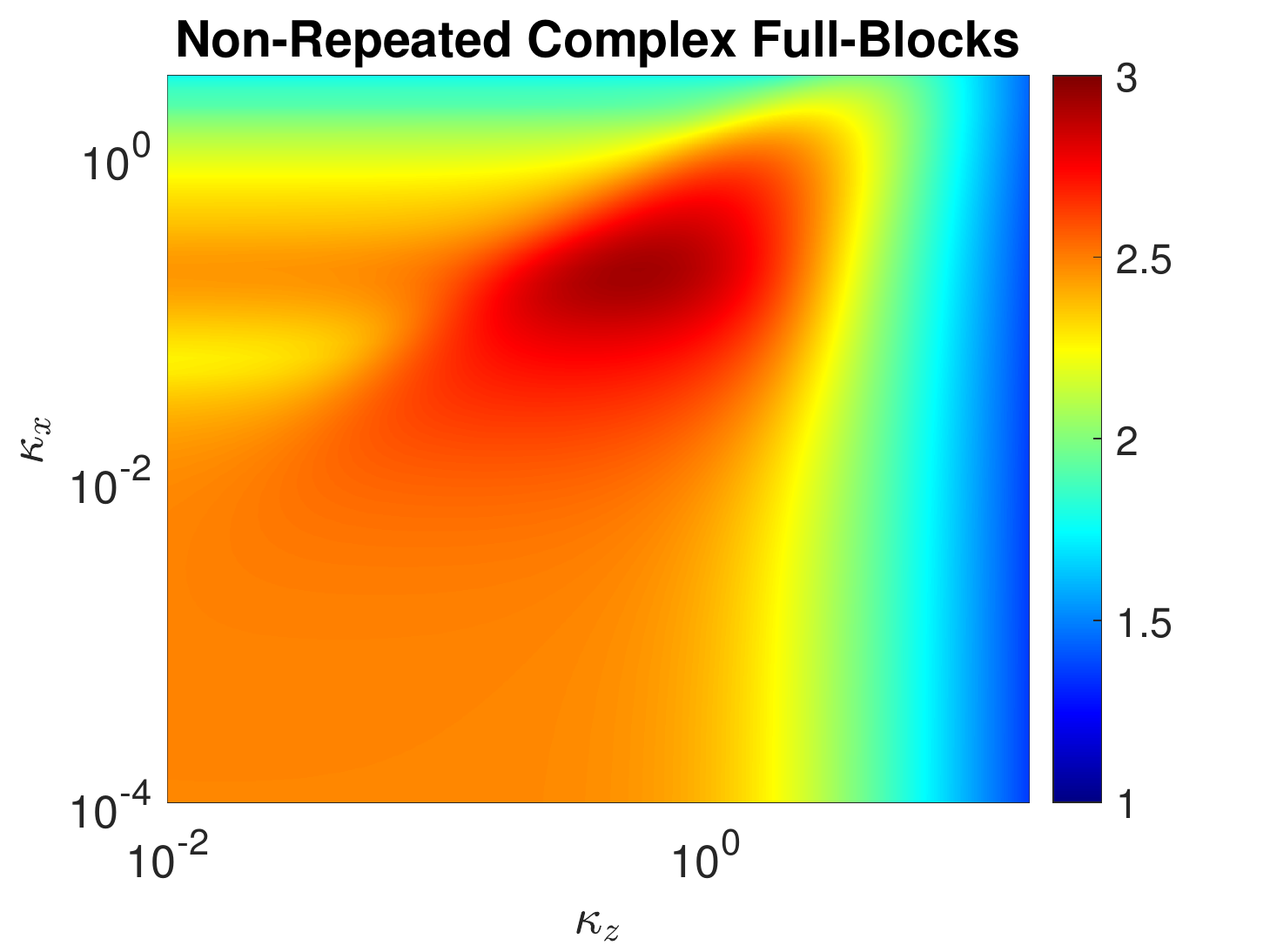} \label{fig:ssv_NFBub}}
     \subfigure[$\log_{10} (\alpha_{\max})$ for $\Delta \in \mathbf{\Delta}_\mathrm{r}$: Algorithm \ref{alg:moc} (Upper Bounds)]{\includegraphics[width = 0.485\textwidth]{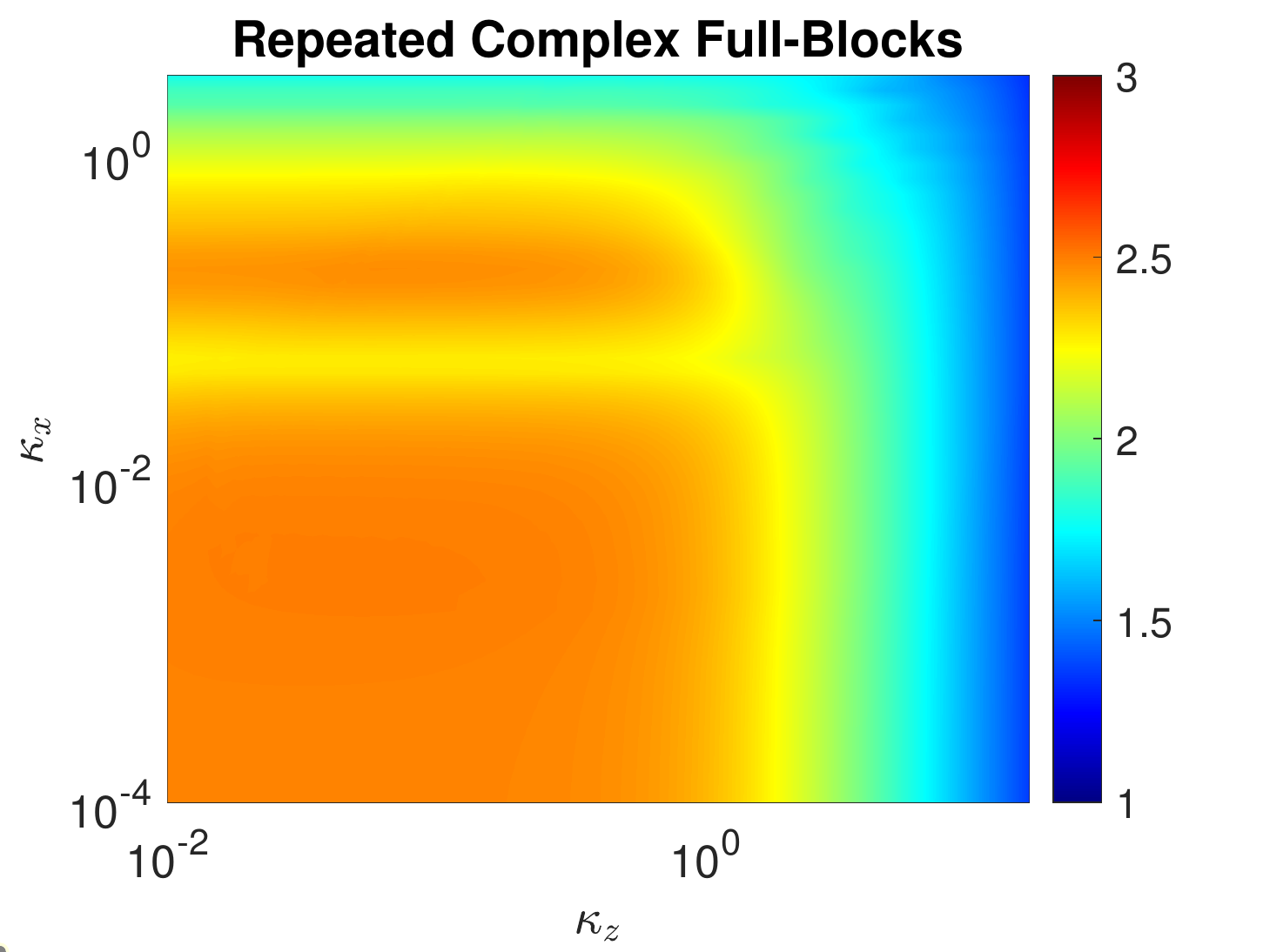} \label{fig:ssv_RFBub}}
     \subfigure[$\log_{10} (\beta_{\max})$ for $\Delta \in \mathbf{\Delta}_\mathrm{nr}$: Algorithm \ref{alg:piter} (Lower Bounds)]{\includegraphics[width = 0.485\textwidth]{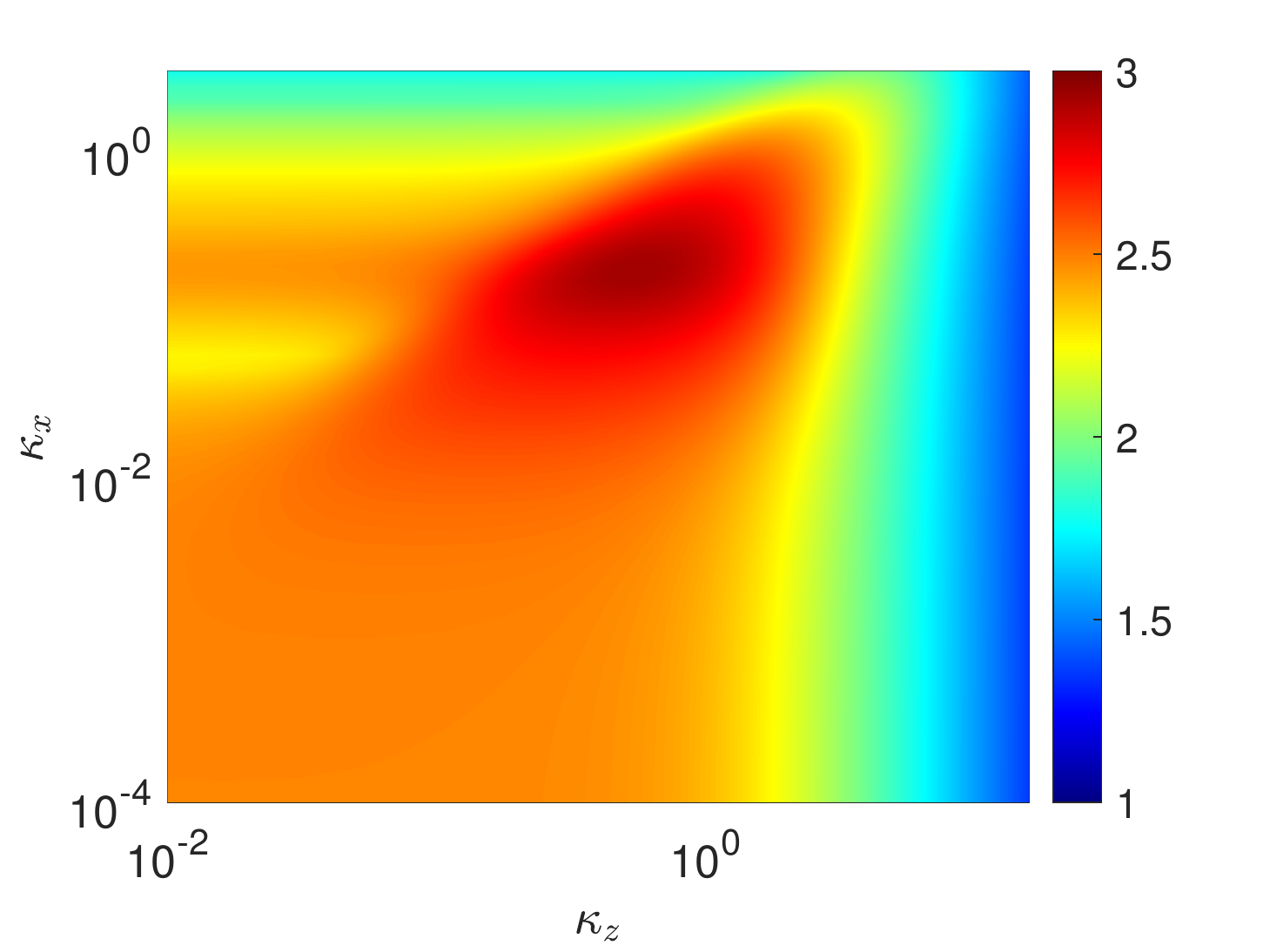} \label{fig:ssv_NFBlb}}
     \subfigure[$\log_{10} (\beta_{\max})$ for $\Delta \in \mathbf{\Delta}_\mathrm{r}$: Algorithm \ref{alg:gen-piter} (Lower Bounds)]{\includegraphics[width = 0.485\textwidth]{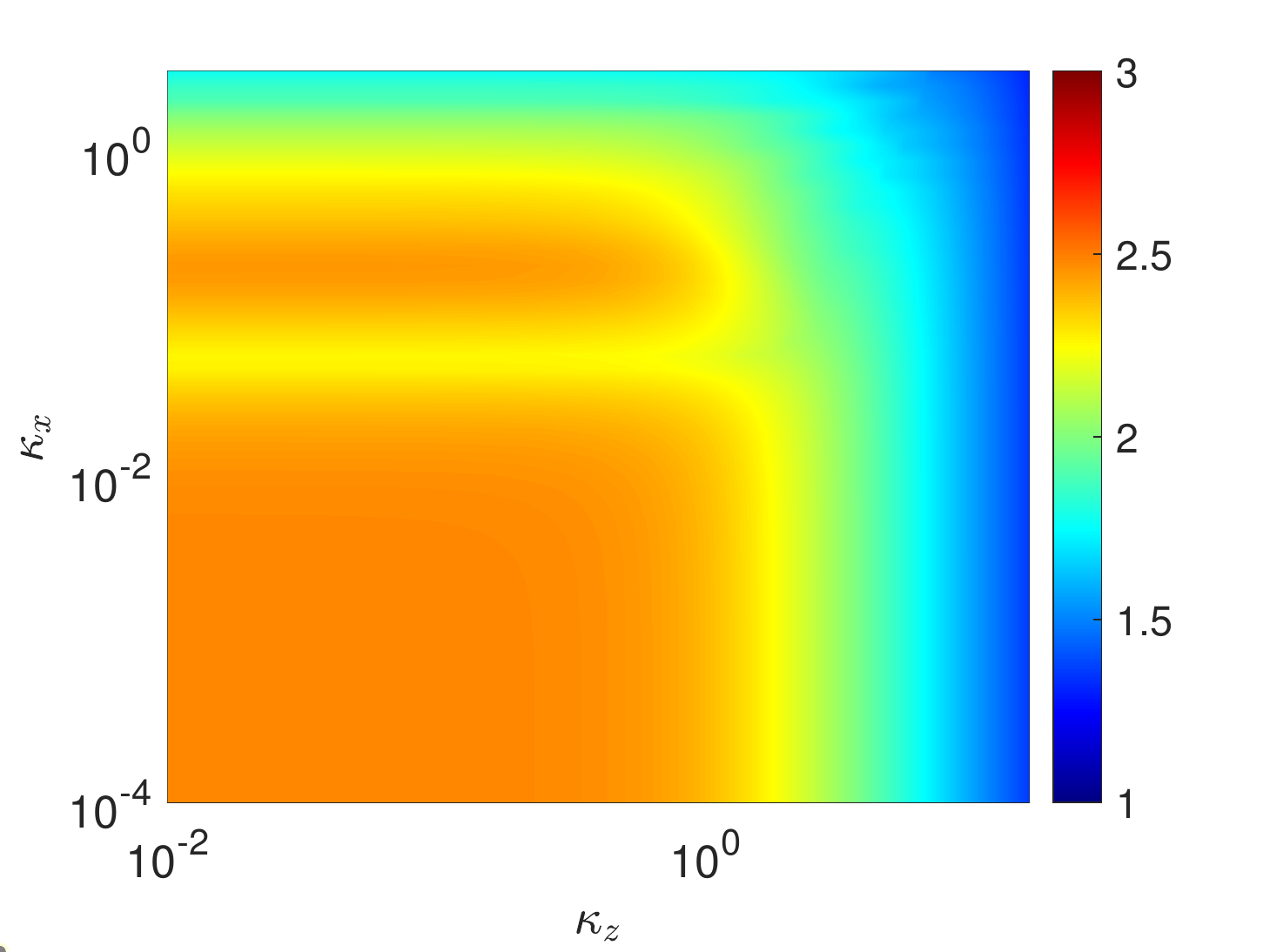} \label{fig:ssv_RFBlb}}
     \end{center}
    % \caption{The upper ($\alpha_{\max}$) and lower ($\beta_{\max}$) bound results over the wavenumber pair $(\kappa_x, \kappa_z)$ grid.}
    \caption{The $\alpha_{\max}$ and $\beta_{\max}$ results over the wavenumber pair $(\kappa_x, \kappa_z)$ grid. The top row % of the figure represents 
    plots represent the % largest 
    upper bounds $\alpha_{\max}$ and the bottom row % of the figure represents 
    plots represent the % largest 
    lower bounds $\beta_{\max}$.}
    \label{fig:ssv_heat_plots}
\end{figure*}
%------------------------------------------------------
% \subsection{Numerical Implementation}
We will use a $50 \times 90 \times 50$ grid of $n_{\kappa_x} \times n_{\kappa_z} \times n_\omega$ to compute the SSV bounds, where $n_{\kappa_x}$, $n_{\kappa_z}$ and $n_{\omega}$ are total grid points for $\kappa_x$, $\kappa_z$ and $\omega$, respectively.
We use % logrithmically
logarithmically spaced values $\kappa_x \in [10^{-4},\; 10^{0.48}] $, $\kappa_z \in [10^{-2},\; 10^{1.2}]$ and $\omega \in [-10^{0.5},\; 10^{0.5}] $ for all the results in this section.
Note that we consider negative temporal frequencies as the system matrices are complex-valued and the corresponding frequency response % for the governing system 
is not symmetric about the $\omega =0$ line.
The state dimension of the system is $s = 30$, and the input and output dimensions are $m = 3s = 90$ and $n = 9s = 270$, respectively.
Then, $m_1 = s = 30$ and $n_1 = 3s = 90$ for $\Delta_1$.
The operating Reynolds number for the system is set to $Re = 358$.
The system is nominally stable, i.e., the eigenvalues of $A(\kappa_x, \kappa_z, Re)$ are in the open left-half plane for the parameter values chosen here.
Algorithm \ref{alg:moc} is initialized % as 
with $R = \text{diag}((d_1^{\star})^2, \ldots, (d_v^{\star})^2) $ using the Osborne's iteration, $p=1.05$, $k_m=500$, $\theta = 10^{-3}$, $\gamma=10^6$ % $r_{\mathrm{cond}} = 10^4$ 
and $\epsilon = 2 \times 10^{-4}$.
%
% Algorithms \ref{alg:piter} and \ref{alg:gen-piter} are initialized by setting $w^{[0]}$ and $b^{[0]}$ to be the left and right unitary vectors associated with largest singular values of $M(\mathrm{i} \omega)$, respectively.
%
Algorithms \ref{alg:piter} and \ref{alg:gen-piter} are initialized by setting $w^{[0]}$ and $b^{[0]}$ to be the right singular vector associated with $\bar{\sigma}\left(D_\mathrm{nr}^\star M(\mathrm{i} \omega) \left(D_\mathrm{nr}^\star\right)^{-1}\right)$, where $D_\mathrm{nr}^\star$ is obtained using the standard Osborne's iterations on $M(\mathrm{i} \omega)$.
Additionally, the total number of iterations given by $k_m$ are set to $60$ for both the power iterations.
Since $M$ in \eqref{eq:transfunc} is a frequency response operator, we will compute the ``best'' upper ($\alpha_{\max}$) and lower ($\beta_{\max}$) bounds at each $(\kappa_x, \kappa_z)$ pair by choosing the maximum $\alpha$ and $\beta$ over a spectrum of frequencies $\omega$.

MATLAB's \texttt{parfor} command is used to compute $\alpha_{\max}$ and $\beta_{\max}$ values using parallel computing for $n_{\kappa_x} \times n_{\kappa_z}$ grid at each $n_\omega$.
%%%%----------------------
%\textcolor{red}{The computations were performed on an ASUS ROG M15 laptop with Intel 2.6 GHz i7-10750H CPU with 6 cores, 16 GB RAM, and an RTX 2070 Max-Q GPU.
%%
%The computation times for Algorithm \ref{alg:moc} and 
%Algorithm \ref{alg:gen-piter}
%were $6$ hours and $41$ minutes, and $56$ minutes, respectively.
%%
%Osborne's iteration and Algorithm \ref{alg:piter} took about 15 minutes each to compute all the $\alpha_{\max}$ and $\beta_{\max}$ values. }
%%%-----------------
The computations were performed on a desktop computer with 3.61 GHz 12-th Gen Intel(R) Core(TM) i7-12700K processor with 12 cores and 16 GB RAM.
The computation times for Algorithm \ref{alg:moc} and 
Algorithm \ref{alg:gen-piter}
were approximately 4 hours and 22 minutes, respectively.
On the other hand, Osborne's iteration and Algorithm \ref{alg:piter} took about 2 minutes and 4 minutes, respectively, to compute all the $\alpha_{\max}$ and $\beta_{\max}$ values. 

%\subsection{Discussion}
The results are depicted in Fig. \ref{fig:ssv_heat_plots}.
Comparing the results shown in Figs. \ref{fig:ssv_RFBub} and \ref{fig:ssv_NFBub}, we deduce that
$\alpha_{\max}$ values computed using Algorithm \ref{alg:moc}
are smaller overall than the $\alpha_{\max}$ values computed using the Osborne's iteration.
The distributions of the $\alpha_{\max}$ values over the wavenumber pair grid are also markedly different. 
There is a prominent peak in Fig. \ref{fig:ssv_NFBub} for the largest $\alpha_{\max}$ value at $\kappa_x = 0.1956$ and $\kappa_z = 0.5778$. This peak is not present in Fig. \ref{fig:ssv_RFBub}. Instead, there are two areas with similar $\alpha_{\max}$ values, which are separated by a narrow `valley' in between.
Therefore, approximating a repeating full-block uncertainty with a non-repeating one in this case not only leads to conservative upper bound estimates, but also results in a local maximum that does not necessarily represent actual system behavior.
A similar argument follows for the lower bounds computed using the two power iteration variants, as shown in Figs. \ref{fig:ssv_NFBlb} and \ref{fig:ssv_RFBlb}.
Additionally, the largest $\alpha_{\max}$ value in Fig. \ref{fig:ssv_RFBub} corresponds to the negative spectrum of temporal frequency grid, which provides further insight into the most sensitive direction for instability of the PCF model in \eqref{eq:transfunc}.

%%%%%%%%%%%
\begin{figure*}[!htb]
    \begin{center}
    \subfigure[$\Delta \in \mathbf{\Delta}_\mathrm{nr}$]{\includegraphics[width = 0.485\textwidth]{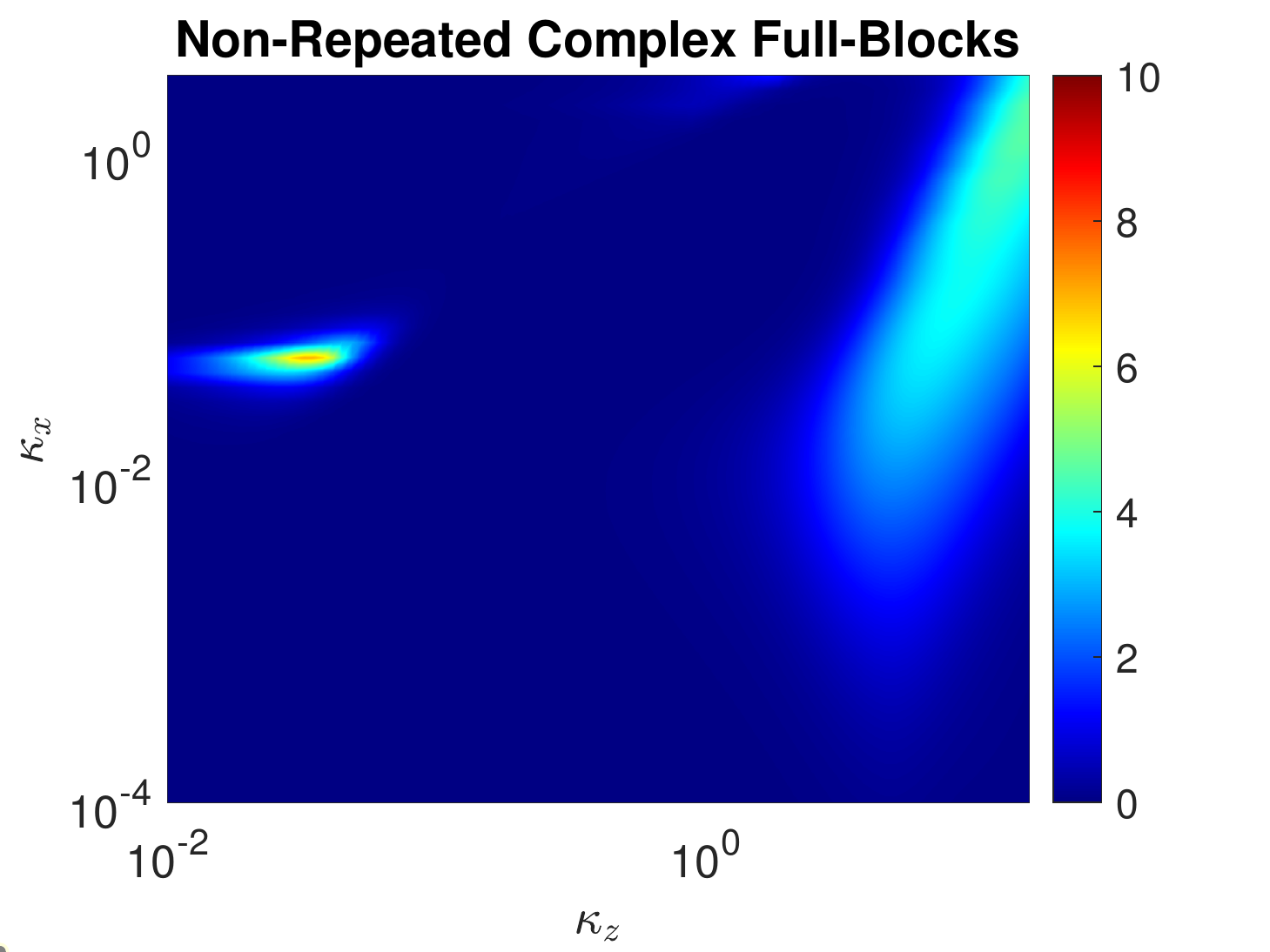} \label{fig:ssv_qualityNFB}}
    \subfigure[$\Delta \in \mathbf{\Delta}_\mathrm{r}$]{\includegraphics[width = 0.485\textwidth]{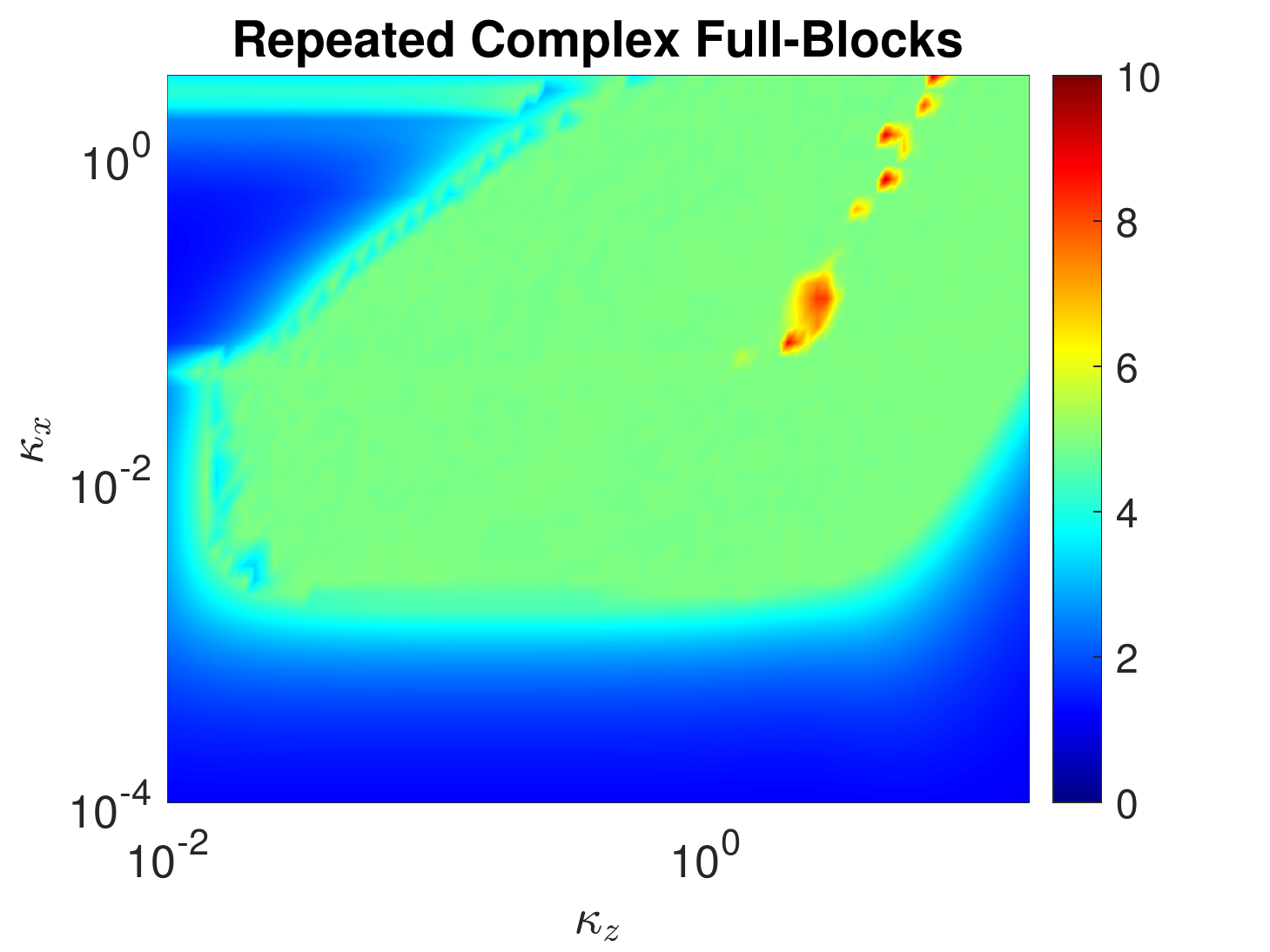} \label{fig:ssv_qualityRFB}}
    \end{center}
    \caption{The percentage difference between $\alpha_{\max}$ and $\beta_{\max}$ values over the wavenumber pair $(\kappa_x, \kappa_z)$ grid.
    The stopping ratio between upper and lower bounds for Algorithm \ref{alg:moc} was set to $1.05$, which means that all the computed upper bounds must be within $5\%$ of the lower bounds.
    Therefore, the majority of percentage differences in (b) are $\leq 5\%$.
    The only upper bounds that failed to achieve the stopping criterion are given by the red hotspots.}
\end{figure*}
%%%%%%%%%%%%%%%%%%%%%%%%%%%%%%%%%%%%%%%%%%%%%%%%%%%%%%%%%%%%%%%%%%%%%%%%
%
The gaps between $\alpha_{\max}$ and $\beta_{\max}$ are shown in Figs. \ref{fig:ssv_qualityNFB} and \ref{fig:ssv_qualityRFB} which indicate that $\beta_{\max}$ values are within $5\%$ of $\alpha_{\max}$ values for approximately $99.8\%$ and $98.9\%$ of wavenumber pairs in Fig. \ref{fig:ssv_qualityNFB} and Fig. \ref{fig:ssv_qualityRFB}, respectively. % for approximately $98\%$ of wavenumber pairs in both sets of bounds. }
This means that the true SSV values lie within a small interval for a large subset of the wavenumber pairs considered for both repeated and non-repeated full-blocks.
A summary of the gaps in both sets of bounds is provided in Table \ref{table:gaps_Couette}.
Although the stopping ratio between upper and lower bounds for Algorithm \ref{alg:moc}  is set to $1.05$, we still end up with a percentage difference greater than $5\%$ at some wavenumber pairs (see Fig. \ref{fig:ssv_qualityRFB}).
However, the maximum gap is $9.33\%$ for only one wavenumber pair and the rest of the wavenumber pairs have an average gap of $6.5\%$ at the hotspots in Fig. \ref{fig:ssv_qualityRFB}. 
%
%The relatively larger gap at these wavenumber pairs is likely due to some numerical issues with either the upper or lower bound algorithms. 
%
The relatively large gap can be attributed to one of the three reasons: (i) The $D$-scale upper bound is not necessarily equal to $\mu$, 
(ii) the upper bound algorithm fails to converge to the optimal $D$-scale, and/or
(iii) the power iteration fails to converge to the true value
of $\mu$.
It is possible that the repeated complex full-blocks are a special
case, where $\mu$ is equal to its corresponding $D$-scale upper bound. 
In
this case, issue (i) would not be the source of the gap.
We will explore this conjecture in future work.

%%%%%%%%%%%%%%%%%%%%%%%
\begin{table}[htb]
\centering
\caption{Summary of the gaps between $\alpha_{\max}$ and $\beta_{\max}$ for the Couette flow model \label{table:gaps_Couette}}%
% \begin{tabular*}{\textheight}{@{\extracolsep\fill}lccc@{\extracolsep\fill}}
\begin{tabular}{ccc}
\toprule
\textbf{Uncertainty Structure} & $\Delta \in \mathbf{\Delta}_\mathrm{nr}$  & $\Delta \in \mathbf{\Delta}_\mathrm{r}$  \\
\midrule
\qquad \textbf{Maximum gap}, $(\kappa_x,\kappa_z)$ & 7.09\%, $(0.055,0.032)$ &  9.33\%,  $(0.692,4.578)$  \\
\midrule
\qquad \textbf{Minimum gap},  $(\kappa_x,\kappa_z)$ & $3.2\times 10^{-5}$\%,  $(0.005,0.011)$ & 1.14\%,  $(10^{-4},10^{1.2})$   \\
\midrule
\qquad \textbf{Average gap} & 0.46\%  & 3.8\% \\
\bottomrule
\end{tabular}
\end{table}
%%%%%%%%%%%%%%%%%%%%%%%%%%%%%%%%%%

%Furthermore, the gap for $\Delta \in \mathbf{\Delta}_\mathrm{nr}$ is at most $11\%$, as shown in Fig. \ref{fig:ssv_qualityNFB}.
%%
%Recall that Osborne's iteration % minimizes 
%utilizes the weaker Frobenius norm bound.
%%
%Thus, in this case, the upper bound computation mainly contributes to this $11\%$ gap between the bounds.
%%
%As previously mentioned in Section \ref{sec:moc}, %we can use
%a variant of Algorithm \ref{alg:moc} can be used to compute $\alpha_{\max}$ for $\Delta \in \mathbf{\Delta}_\mathrm{nr}$. Doing so reduces the hotspots in Fig. \ref{fig:ssv_qualityNFB} as the corresponding $\alpha_{\max}$ values are lower (at most $8.5\%$ less) than the $\alpha_{\max}$ values computed using Osborne's iteration at those wavenumber pairs. Also, the gap remains within $1\%$ everywhere else.
%%
%It is noteworthy that the total computation time for Osborne's iteration is about 15 minutes, whereas the Algorithm \ref{alg:moc} variant for $\Delta \in \mathbf{\Delta}_\mathrm{nr}$ takes about $5$ hours and $12$ minutes to compute all the $\alpha_{\max}$ values. Therefore, the improvements offered by the Algorithm \ref{alg:moc} variant come at a significant increase in computational time. 

%%%%%%%%%%%
\begin{figure*}[!htb]
    \begin{center}
    \subfigure[$(\kappa_x,\kappa_z)=(0.055,0.032)$]{\includegraphics[width = 0.485\textwidth]{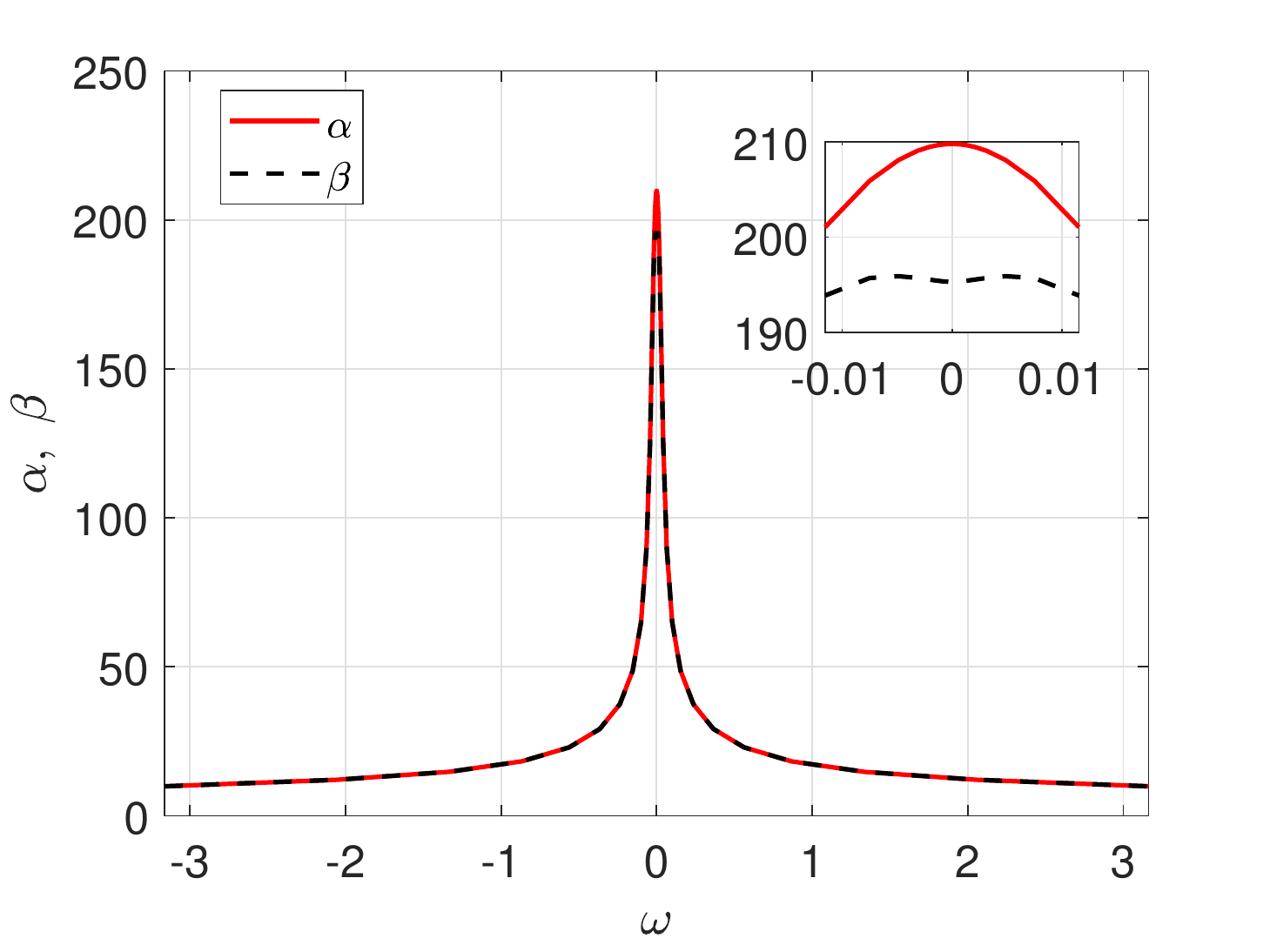} \label{fig:both_bounds_Couette_NFB_1}}
    \subfigure[$(\kappa_x,\kappa_z)=(0.005,0.011)$]{\includegraphics[width = 0.485\textwidth]{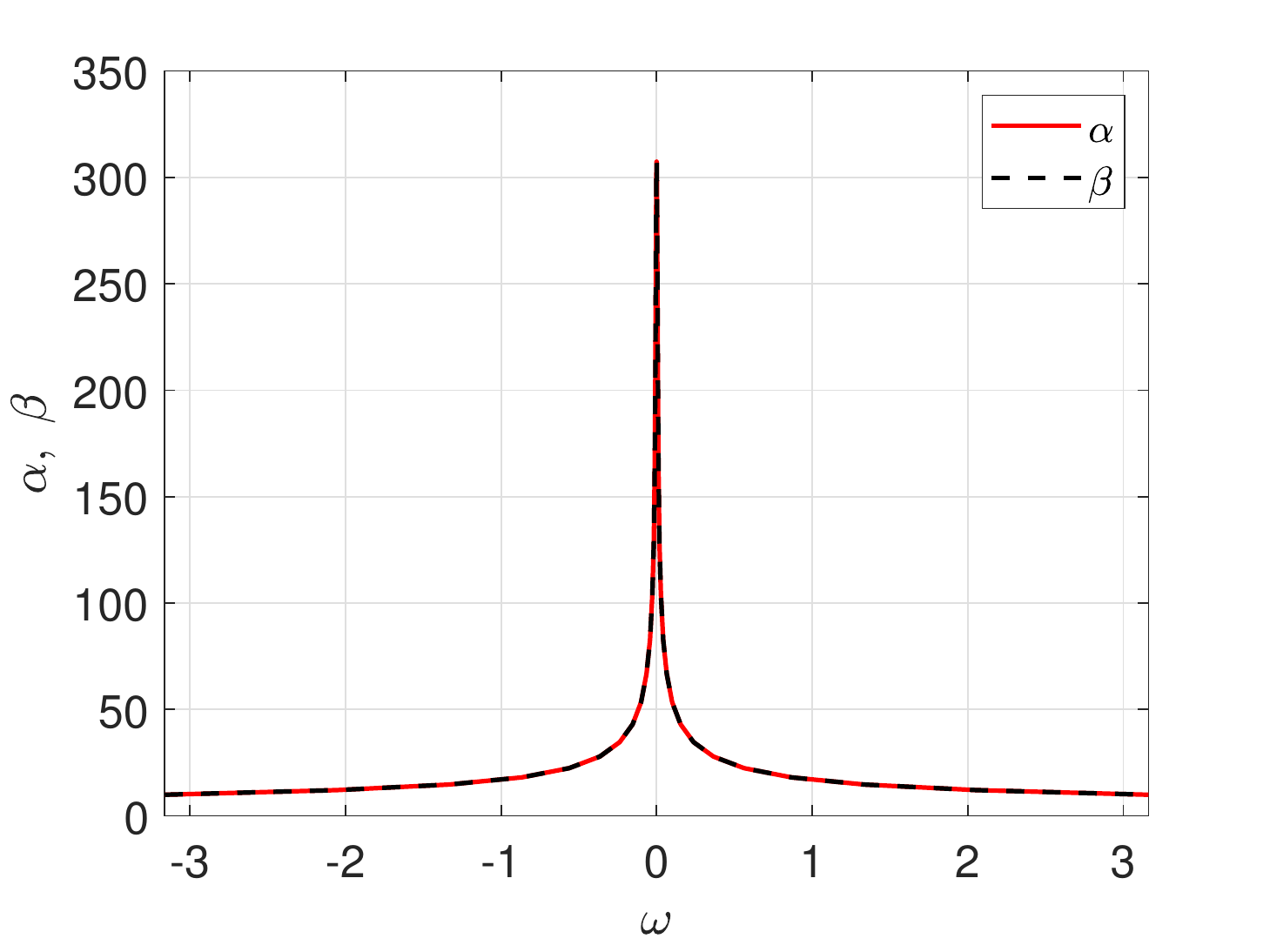} \label{fig:both_bounds_Couette_NFB_2}}
    \end{center}
    \caption{The $\alpha$ and $\beta$ results over the temporal frequency ($\omega$) grid for $\Delta \in \mathbf{\Delta}_\mathrm{nr}$. The (a) and (b) correspond to wavenumber pairs where the gap between $\alpha_{\max}$ and $\beta_{\max}$ are the largest and smallest, respectively. }
    \label{fig:Couette_bounds_vs_frequency_NFB}
\end{figure*}
%%%%%%%%%%%%%%%%%%%%%

\begin{figure*}[!htb]
    \begin{center}
    \subfigure[$(\kappa_x,\kappa_z)=(0.692,4.578)$]{\includegraphics[width = 0.485\textwidth]{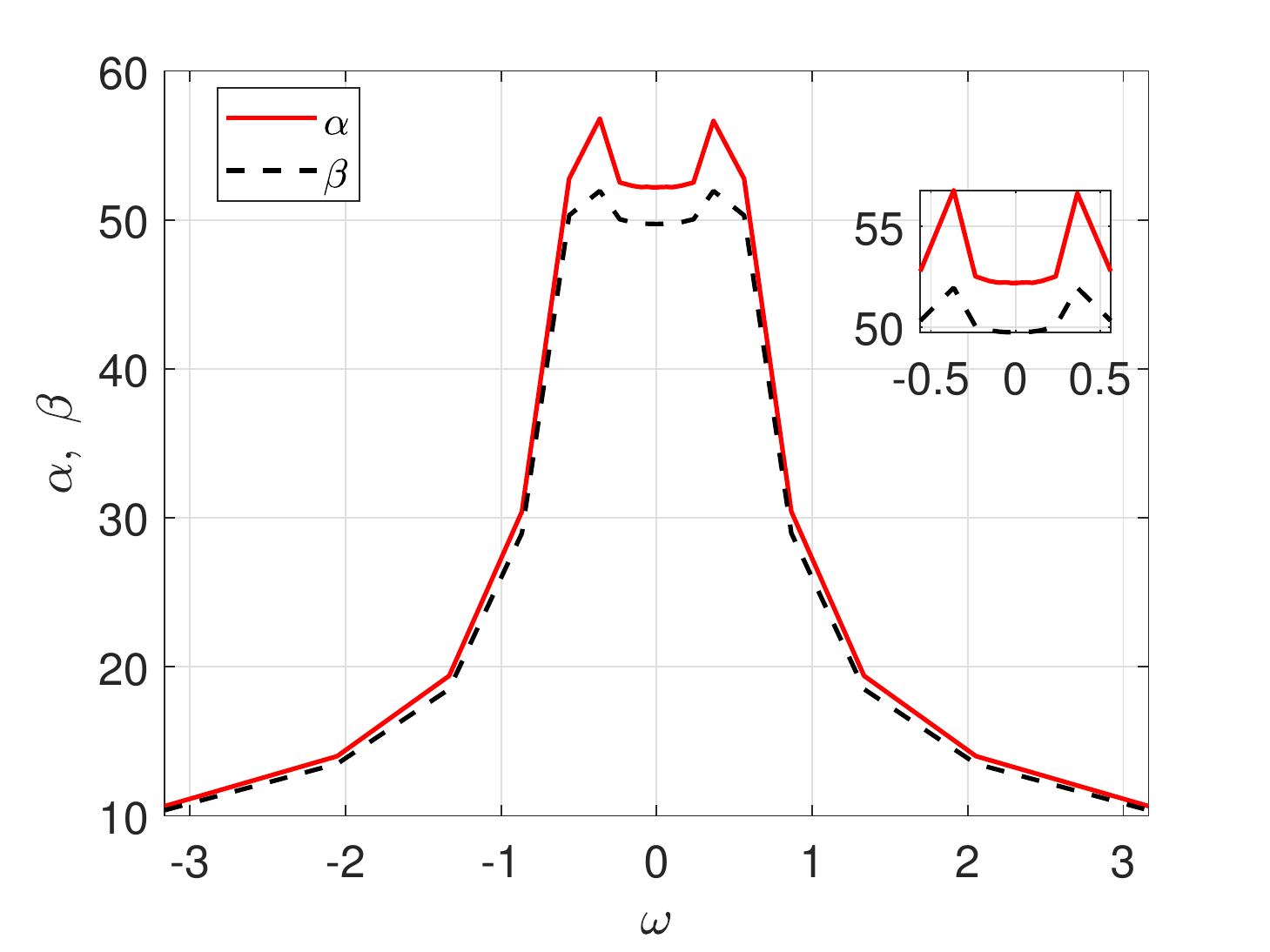} \label{fig:both_bounds_Couette_RFB_1}}
    \subfigure[$(\kappa_x,\kappa_z)=(10^{-4},10^{1.2})$]{\includegraphics[width = 0.485\textwidth]{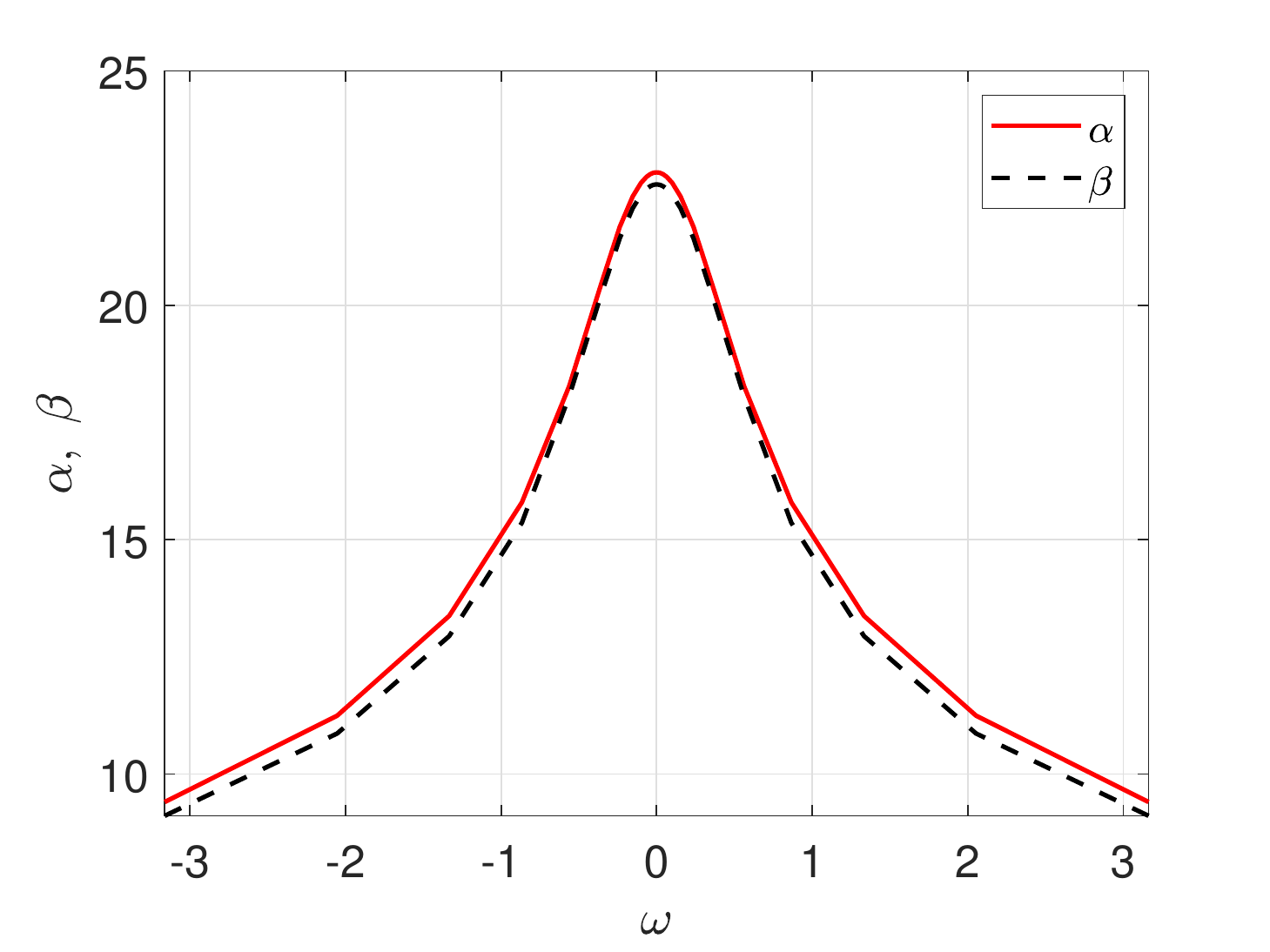} \label{fig:both_bounds_Couette_RFB_2}}
    \end{center}
    \caption{The $\alpha$ and $\beta$ results over the temporal frequency ($\omega$) grid for $\Delta \in \mathbf{\Delta}_\mathrm{r}$. The (a) and (b) correspond to wavenumber pairs where the gap between $\alpha_{\max}$ and $\beta_{\max}$ are the largest and smallest, respectively. }
    \label{fig:Couette_bounds_vs_frequency_RFB}
\end{figure*}

To further investigate the bounds, we plot $\alpha$ and $\beta$ over the temporal frequency at chosen wavenumber pairs $(\kappa_x, \kappa_z)$. 
These results for the non-repeated and repeated full-blocks are shown in Figs. \ref{fig:Couette_bounds_vs_frequency_NFB} and \ref{fig:Couette_bounds_vs_frequency_RFB}, respectively. 
The wavenumber pairs chosen are the ones corresponding to the largest and smallest gap between $\alpha_{\max}$ and $\beta_{\max}$ over the wavenumber pair grid.
The result in Fig. \ref{fig:both_bounds_Couette_NFB_1} showcases the bounds for the non-repeated full blocks at $(\kappa_x,\kappa_z)=(0.055,0.032)$, where the gap between $\alpha_{\max}$ and $\beta_{\max}$ is the largest at 7.09\%. 
The zoomed-in plot in Fig. \ref{fig:both_bounds_Couette_NFB_1} highlights a single global peak in $\alpha$ at $\omega \approx 0$, while there are two local peaks in $\beta$, located almost symmetrically about the $\omega=0$ line at $\omega = \pm 0.005$.
%
% This plot also reveals that the largest gap occurs at $\omega \approx 0$ and the gaps are smaller for larger values of $\omega$.
%
In the case of smallest gap between $\alpha_{\max}$ and $\beta_{\max}$ for the non-repeated full-blocks, which occurs at $(\kappa_x,\kappa_z)=(0.005,0.011)$, the bounds are virtually identical (see Fig. \ref{fig:both_bounds_Couette_NFB_2}). 
On the other hand, both the bounds are qualitatively similar for the repeated full-blocks case, as shown in Fig. \ref{fig:Couette_bounds_vs_frequency_RFB}. 
The largest gap between $\alpha_{\max}$ and $\beta_{\max}$ in this case occurs at $(\kappa_x,\kappa_z)=(0.692,4.578)$, and both $\alpha$ and $\beta$ have two local peaks that occur at $\omega = \pm 0.365$ (see Fig. \ref{fig:both_bounds_Couette_RFB_1}). 
Although these peaks are symmetric about the $\omega=0$ line, the peak $\alpha$ values in Fig. \ref{fig:both_bounds_Couette_RFB_1} at $\omega=-0.365$ and $\omega=0.365$ are 56.799 and 56.651, respectively. 

%%%%%%%%%%%%%%%%%%%
%%%%%%%%%%%

%%%%%%%%%%%%%%%%%%%%%%%%%%%%%%%%%%%%%%%%%%%%%%%%%%%%%%%%%%%%%%%%%%%%%%%%%%%%%%%%%%%%%%%
%%%%%%%%%%%%%%%%%%%%%%%%%%%%%%%%%%%%%%%%%%%%%%%%%%%%%%%%%%%%%%%%%%%%%%%%%%%%%%%%%%%%%%
%%%%%%%%%%%
%
\subsection{Simple Academic Example} \label{Sec:toy problem}
\begin{figure*}[!htb]
    \begin{center}
    \subfigure[$\Delta \in \mathbf{\Delta}_\mathrm{nr}$]{\includegraphics[width = 0.485\textwidth]{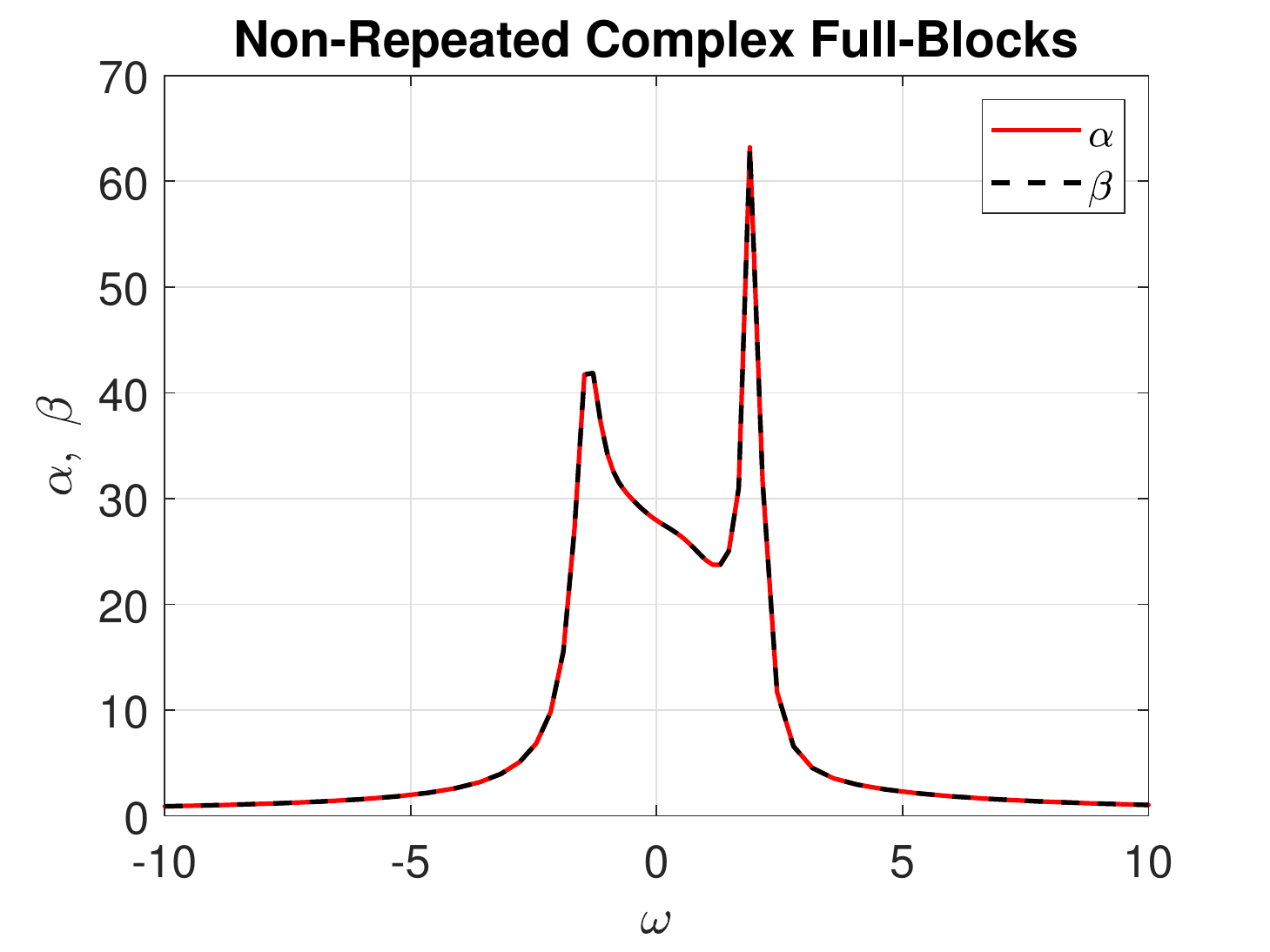} \label{fig:toy_NFB}}
    \subfigure[$\Delta \in \mathbf{\Delta}_\mathrm{r}$]{\includegraphics[width = 0.485\textwidth]{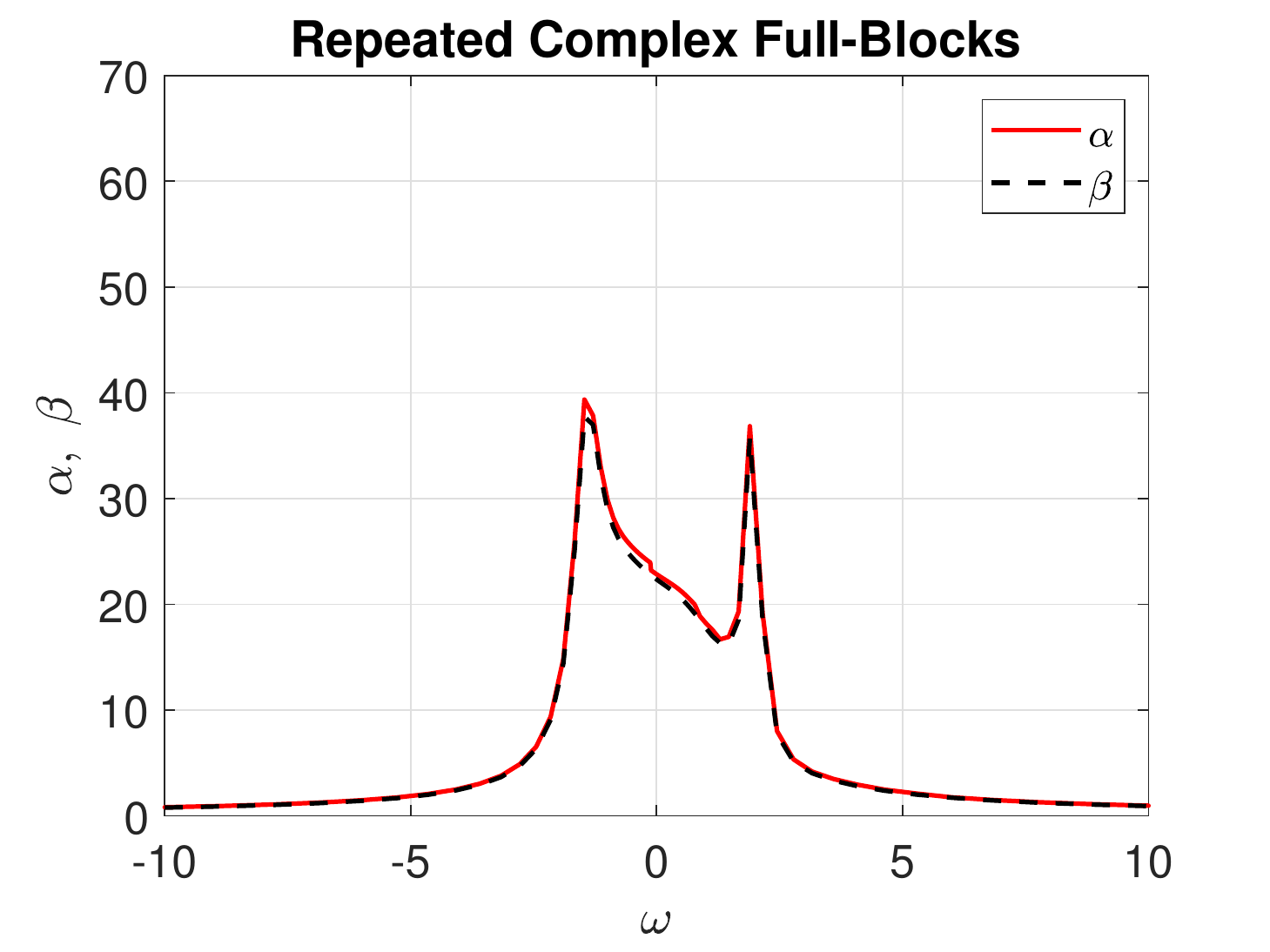} \label{fig:toy_RFB}}
    \end{center}
    \caption{The $\alpha$ and $\beta$ results over the temporal frequency ($\omega$) grid. Although we consider $\omega \in [-10^{1.5}, 10^{1.5}]$, the results are shown for $\omega \in [-10,10]$ to better highlight the local behavior of the bounds. }
    \label{fig:toy_problem}
\end{figure*}

We now demonstrate the proposed algorithms on a % multiple-input multiple-output (MIMO)
MIMO LTI system with the frequency response matrix given by $M = C (\mathrm{i}\omega I_{4} - A)^{-1}B,$ where $\omega$ is the temporal frequency and the randomly generated state-space matrices $A, B, C \in \mathbb{C}^{4 \times 4}$ are as follows:
\begin{align*}
A = \begin{bmatrix}
    0.720 -\mathrm{i} 0.663 & -0.602 - \mathrm{i} 0.684 & -1.937 - \mathrm{i} 0.792 & -1.021 -\mathrm{i} 0.153 \cr
    0.059 - \mathrm{i} 1.875 & -1.103 + \mathrm{i} 0.350 & -0.728 + \mathrm{i} 0.164 & -0.135 + \mathrm{i} 2.021 \cr
    0.071 + \mathrm{i} 0.114 &  0.948 + \mathrm{i} 0.237 & -1.493 + \mathrm{i} 0.491 &  1.486 - \mathrm{i} 0.025 \cr
   -0.647 - \mathrm{i} 0.260 & -0.272 + \mathrm{i} 0.829 & -0.709 + \mathrm{i} 0.908 & -0.506 + \mathrm{i} 0.276
  \end{bmatrix}, \\
%%%%%%%%%%%%%
B =  \begin{bmatrix}
    0.738 - \mathrm{i} 0.773 &  1.271 + \mathrm{i} 0.118 &  1.152 + \mathrm{i} 0.494 & -0.764 - \mathrm{i} 0.400 \cr
   -0.166 + \mathrm{i} 0.896 &  0.504 + \mathrm{i} 1.761 &  0.291 - \mathrm{i} 0.516 &  0.425 - \mathrm{i} 0.028 \cr
   -1.103 + \mathrm{i} 0.449 & -1.408 -\mathrm{i} 0.195 &  0.067 - \mathrm{i} 1.287 & -0.595 + \mathrm{i} 0.316 \cr
    1.308 - \mathrm{i} 0.744 &  0.358 + \mathrm{i} 0.728 & -0.174 + \mathrm{i} 0.665 & -1.489 - \mathrm{i} 0.094
  \end{bmatrix}, \\
  %%%%%%%%%%%%%%
C = \begin{bmatrix}
    0.255 + \mathrm{i} 0.101 &  1.681 + \mathrm{i} 0.048 & -0.386 - \mathrm{i} 0.051 &  0.633 - \mathrm{i} 0.874 \cr
   -1.827 + \mathrm{i} 1.132 & -0.267 - \mathrm{i} 0.846 & -0.863 + \mathrm{i} 0.840 &  0.244 + \mathrm{i} 1.447 \cr
    1.877 + \mathrm{i} 0.179 & -1.124 + \mathrm{i} 0.752 &  1.014 + \mathrm{i} 0.731 & -1.502 + \mathrm{i} 0.431 \cr
   -0.803 + \mathrm{i} 1.056 &  0.002 - \mathrm{i} 0.284 &  1.029 - \mathrm{i} 0.801 & -0.444 + \mathrm{i} 0.543
  \end{bmatrix}.
\end{align*}
The nominal system is stable as all the eigenvalues of $A$ are in the open left-half plane.
The uncertainty for this model is chosen as $\Delta = I_2 \otimes \Delta_1$ with $\Delta_1 \in \mathbb{C}^{2 \times 2}$.
Numerical implementation of the algorithms are as described in Section \ref{Sec:Example Model-1}.
%
% Parameters related to the algorithms are same as described in Section \ref{Sec:Example Model-1}. 
%
We take 200 logarithmically spaced points for $\omega \in [-10^{1.5}, 10^{1.5}]$. 
The results for $\alpha$ and $\beta$ for both the non-repeated and repeated cases are shown in Fig. \ref{fig:toy_problem}. 
In terms of qualitative similarities, there are two peaks--one for $\omega<0$ and the other for $\omega >0$-- in each bound in both the cases, and the bounds are not symmetric about the $\omega=0$ line. 
However, approximating the repeated full-block structure with a non-repeated one leads to very conservative bounds at some temporal frequencies. % the bounds obtained by treating the repeated structure as non-repeated are very conservative. %Comparing the bounds in Fig. \ref{fig:toy_NFB} with the ones in Fig. \ref{fig:toy_RFB}, the conservatism in the case of approximating the repeated full-blocks as non-repeated is clear. 
For example, the $\alpha$ value in Fig. \ref{fig:toy_NFB} is approximately 1.7 times that of the $\alpha$ value in Fig. \ref{fig:toy_RFB} at $\omega=1.896$. 
A similar set of comments applies to the lower bounds $\beta$ at $\omega=1.896$. 
This means that the true value of $\mu$ at this frequency in the non-repeated case is roughly 1.7 times that of the repeated case.
It is also noteworthy that the global peaks of the bounds in Fig. \ref{fig:toy_NFB} are at $\omega > 0$, whereas the global peaks in Fig. \ref{fig:toy_RFB} are at $\omega < 0$. 
Therefore, similar to the fluid-flow example, neglecting the repeated structure of the uncertainty can not only lead to conservative bounds, which translates into conservative stability-margin estimates, but also might lead to inaccurate conclusions about the temporal behavior of the system.
%
%%%%%%%%%%%%%%%
%%%%%%%%%%%%%%%%%%%%%%%%%%%
%%%%%%%%%%%%%%%%%%%%%%%%%%%%%%%%%%%%%%%%%%%%%%%%%%%
%%%%%%%%%%%%%%%%%%%%%%%%%%%%%%%%%%%%%%%%%%%%%%%%%%%%%%%%%%%%%%%%%%%%%%%%%%%%%%%%%%%%%%%%%%%%%%%%%%%%%%%%%%%%%%%%%%%%%%%%%%%%%%%%%%%%%%%%%%%%%%%%%%%%%%%%%%%%%%
\section{Conclusions} \label{sec:conclusions}
We proposed two algorithms for computing upper and lower bounds of structured singular value for repeated complex full-block uncertainty. 
Such uncertainty structures naturally arise in models of fluid flows and other convective systems.
The proposed algorithms yield bounds that are less conservative as compared to the algorithms that ignore the repeated full-block structure, e.g., Osborne's iteration for non-repeated full-blocks.
%
%\textcolor{blue}{Therefore, ignoring the repeated structure of the uncertainty can lead to conservative stability margin estimates for the system.}
%
Thus, properly accounting for the repeated block structure can improve stability-margin estimates and also enable one to draw more representative conclusions regarding the temporal behavior of the system.
These points were demonstrated on an example of incompressible plane Couette flow and an academic example.
%These algorithms can be applied to investigate stability of fluid flows and other convective systems. 
%
Furthermore, our future work will involve investigating the gap between the $\mu$ %structured singular value 
and the convex (or $D$-scale) upper bound for a single repeated full-block.
This particular case is of interest due to the fact that $\mu$ is equal to its upper bound for a single full block and also for a single repeated complex scalar.
%
% \textcolor{blue}{Based on the tightness of the bounds presented in this work, we conjecture that $\mu$ is equal to its upper bound for a single repeated full block.
% %
% This conjecture is further made plausible by the fact that $\mu$ is equal to its upper bound for a single full block and also for a single repeated complex scalar.
% %, so , however, not necessarily true.
% %
% Therefore, our future work involves investigating the gap between the structured singular value and the convex (or $D$-scale) upper bound for a single repeated full block to establish the legitimacy of this conjecture.}

%-------------------------------------------------------
\section*{Acknowledgements}
This material is based upon work supported by the Air Force Office of Scientific Research under award number FA9550-21-1-0106, the Army Research Office under award number W911NF-20-1-0156, the National Science Foundation under award number CBET-1943988, and the Office of Naval Research under award number N00014-22-1-2029. 
%-------------------------------------------
\bibliography{ref}

\begin{thebibliography}{10}
\providecommand \doibase [0]{http://dx.doi.org/}%

\bibitem{doyle1982analysis}
Doyle J. Analysis of feedback systems with structured uncertainties. In: IEE
  Proceedings D (Control Theory and Applications). ; 1982\string: 242--250

\bibitem{packard1993complex}
Packard A, Doyle J. The complex structured singular value. {\it Automatica}
  1993\string; 29(1)\string: 71--109.
\newblock \href {\doibase 10.1016/0005-1098(93)90175-S} {doi:
  10.1016/0005-1098(93)90175-S}

\bibitem{safonov82}
Safonov MG. Stability margins of diagonally perturbed multivariable feedback
  systems. In: 20th IEEE Conference on Decision and Control including the
  Symposium on Adaptive Processes. ; 1981\string: 1472-1478

\bibitem{zhou1996robust}
Zhou K, Doyle J, Glover K. {\it Robust and Optimal Control}.
\newblock Feher/Prentice Hall Digital andPrentice Hall .
\newblock 1996.

\bibitem{dullerud13}
Dullerud GE, Paganini F. {\it A course in robust control theory: a convex
  approach}. 36.
\newblock Springer Science \& Business Media .
\newblock 2013.

\bibitem{braatz1994computational}
Braatz RP, Young PM, Doyle JC, Morari M. Computational complexity of $\mu$
  calculation. {\it IEEE Transactions on Automatic Control} 1994\string;
  39(5)\string: 1000--1002.
\newblock \href {\doibase 10.1109/9.284879} {doi: 10.1109/9.284879}

\bibitem{demmel92}
Demmel J. The Componentwise Distance to the Nearest Singular Matrix. {\it SIAM
  J. Matrix Anal. Appl.} 1992\string; 13(1)\string: 10-19.
\newblock \href {\doibase 10.1137/0613003} {doi: 10.1137/0613003}

\bibitem{young90}
Young P, Doyle J. Computation of mu with real and complex uncertainties. In:
  29th IEEE Conference on Decision and Control. ; 1990\string: 1230-1235 vol.3

\bibitem{young1992}
Young PM, Newlin MP, Doyle JC. Practical computation of the mixed $\mu$
  problem. In: American Control Conference. ; 1992\string: 2190-2194

\bibitem{Troeng2021}
Troeng O. Five-Full-Block Structured Singular Values of Real Matrices Equal
  Their Upper Bounds. {\it IEEE Control Systems Letters} 2021\string;
  5(2)\string: 583-586.
\newblock \href {\doibase 10.1109/LCSYS.2020.3004297} {doi:
  10.1109/LCSYS.2020.3004297}

\bibitem{colombino2016}
Colombino M, Smith RS. A Convex Characterization of Robust Stability for
  Positive and Positively Dominated Linear Systems. {\it IEEE Transactions on
  Automatic Control} 2016\string; 61(7)\string: 1965-1971.
\newblock \href {\doibase 10.1109/TAC.2015.2480549} {doi:
  10.1109/TAC.2015.2480549}

\bibitem{Fan1991}
Fan M, Tits A, Doyle J. Robustness in the presence of mixed parametric
  uncertainty and unmodeled dynamics. {\it IEEE Transactions on Automatic
  Control} 1991\string; 36(1)\string: 25-38.
\newblock \href {\doibase 10.1109/9.62265} {doi: 10.1109/9.62265}

\bibitem{packard1988}
Packard A, Fan M, Doyle J. A power method for the structured singular value.
  In: 27th IEEE Conference on Decision and Control. ; 1988\string: 2132-2137
  vol.3.

\bibitem{liu21}
Liu C, Gayme DF. Structured input–output analysis of transitional
  wall-bounded flows. {\it Journal of Fluid Mechanics} 2021\string; 927.
\newblock \href {\doibase 10.1017/jfm.2021.762} {doi: 10.1017/jfm.2021.762}

\bibitem{liu2022strat}
Liu C, Colm-cille PC, Gayme DF. Structured input--output analysis of stably
  stratified plane Couette flow. {\it Journal of Fluid Mechanics} 2022\string;
  948\string: A10.
\newblock \href {\doibase 10.1017/jfm.2022.648} {doi: 10.1017/jfm.2022.648}

\bibitem{liu2023structured}
Liu C, Shuai Y, Rath A, Gayme DF. A structured input-output approach to
  characterizing optimal perturbations in wall-bounded shear flows. In:
  American Control Conference. ; 2023\string: 2319-2325

\bibitem{Bhattacharjee_et_al_2023}
Bhattacharjee D, Mushtaq T, Seiler PJ, Hemati M. Structured Input-Output
  Analysis of Compressible Plane Couette Flow. In: AIAA SCITECH FORUM. ;
  2023\string: 1984

\bibitem{balas2007robust}
Balas G, Chiang R, Packard A, Safonov M. Robust control toolbox user’s guide.
  {\it The Math Works, Inc., Tech. Rep} 2007.

\bibitem{boyd1993method}
Boyd S, {El Ghaoui} L. Method of centers for minimizing generalized
  eigenvalues. {\it Linear Algebra and its Applications} 1993\string;
  188-189\string: 63-111.
\newblock \href {\doibase https://doi.org/10.1016/0024-3795(93)90465-Z} {doi:
  https://doi.org/10.1016/0024-3795(93)90465-Z}

\bibitem{boyd1994linear}
Boyd S, El~Ghaoui L, Feron E, Balakrishnan V. {\it Linear Matrix Inequalities
  in System and Control Theory}.
\newblock Society for Industrial and Applied Mathematics .
\newblock 1994

\bibitem{safonov1980stability}
Safonov MG. {\it Stability and Robustness of Multivariable Feedback Systems}.
\newblock MIT press .
\newblock 1980.

\bibitem{nesterov1995interior}
Nesterov YE, Nemirovskii A. An interior-point method for generalized
  linear-fractional programming. {\it Mathematical Programming} 1995\string;
  69(1)\string: 177--204.
\newblock \href {\doibase 10.1007/BF01585557} {doi: 10.1007/BF01585557}

\bibitem{mehrotra1992implementation}
Mehrotra S. On the Implementation of a Primal-Dual Interior Point Method. {\it
  SIAM Journal on Optimization} 1992\string; 2(4)\string: 575-601.
\newblock \href {\doibase 10.1137/0802028} {doi: 10.1137/0802028}

\bibitem{osborne_iter}
Osborne EE. On Pre-Conditioning of Matrices. {\it Journal of the ACM}
  1960\string; 7(4)\string: 338–345.
\newblock \href {\doibase 10.1145/321043.321048} {doi: 10.1145/321043.321048}

\bibitem{McKeonJFM2010}
McKeob BJ, Sharma AS. A critical-layer framework for turbulent pipe flow. {\it
  Journal of Fluid Mechanics} 2010\string; 658\string: 336–382.
\newblock \href {\doibase 10.1017/S002211201000176X} {doi:
  10.1017/S002211201000176X}

\bibitem{chavarin2020resolvent}
Chavarin A, Luhar M. Resolvent Analysis for Turbulent Channel Flow with
  Riblets. {\it AIAA Journal} 2020\string; 58(2)\string: 589-599.
\newblock \href {\doibase 10.2514/1.J058205} {doi: 10.2514/1.J058205}

\bibitem{liu20}
Liu C, Gayme DF. Input-output inspired method for permissible perturbation
  amplitude of transitional wall-bounded shear flows. {\it Phys. Rev. E}
  2020\string; 102\string: 063108.
\newblock \href {\doibase 10.1103/PhysRevE.102.063108} {doi:
  10.1103/PhysRevE.102.063108}

\bibitem{jovanovic2005componentwise}
Jovanovi{\'c} MR, Bamieh B. Componentwise energy amplification in channel
  flows. {\it Journal of Fluid Mechanics} 2005\string; 534\string: 145-183.
\newblock \href {\doibase 10.1017/S0022112005004295} {doi:
  10.1017/S0022112005004295}

\bibitem{Boyd2004}
Boyd S, Vandenberghe L. {\it {Convex Optimization}}.
\newblock Cambridge University Press .
\newblock 2004.

\end{thebibliography}
\appendix
\section{Generalized Osborne} \label{sec:appendix}
In this section, we will describe a fast algorithm for $\Delta \in \mathbf{\Delta}_\mathrm{r}$ and $M \in \mathbb{C}^{m \times m}$.
The standard Osborne iteration cannot be used for $\mathbf{\Delta}_\mathrm{r}$ as $D \in \mathbf{D}_\mathrm{r}$ contains off-diagonal entries.
This section describes our generalization of Osborne's method (GenOsborne) to handle the matrix scales in \eqref{eq:scaling_struct_2}.
The proposed GenOsborne algorithm is an iteration that solves the following minimization problem:
\begin{align}
    \begin{aligned}
        & \min_{D\in \mathbf{D}_\mathrm{r}}  \|D MD^{-1}\|^2_F 
    \end{aligned}
    \label{eq:min_prob_2}
\end{align} 
where $\mathbf{D}_\mathrm{r}$ is defined in \eqref{eq:scaling_struct_2}.
To simplify the calculations, we use the square of Frobenius norm in \eqref{eq:min_prob_2}.
Let $s_{ij}$ denote the $(i,j)$ entry of $S$ in \eqref{eq:scaling_struct_2}.
The Frobenius norm in \eqref{eq:min_prob_2} yields a cumbersome expression that has various $s_{ij} \in \mathbb{C}$ entries multiplying each other.
Thus, it is difficult to minimize the function for each $s_{ij}$, since each of the scalings are coupled together.
To avoid this issue, we iteratively optimize over a single off-diagonal entry and then couple it, similar to the Osborne's iteration.
Thus, we first use the standard Osborne's iterations to calculate the optimal diagonal scalings $s_{i}^\star$ and then use an iterative approach to
optimize a single off-diagonal term $s_{ij} \in \mathbb{C}$ at each iteration and iterate over all possible pairs of $(i,j)$, where $i \neq j$.
We denote the matrices with a single off-diagonal entry $s_{ij} \in \mathbb{C}$ as 
$D_{ij} = S_{ij}\otimes I_{m_1}$, where $S_{ij}$ has ones along the diagonal, $s_{ij}$ in the $(i,j)$ entry and zero everywhere else.

Let $M^{[k]}$ be the scaled matrix at step $k$ of the generalized iteration and $s_{ij}\in \mathbb{C}$ be the off-diagonal scaling to be optimized. Then, the objective function is:
\begin{align}
    \label{eq:general_fro}
    f_1(s_{ij}) &= \| D_{ij} M^{[k]} D_{ij}^{-1} \|_F^2 \\
    \nonumber
    &= c_0 + \text{conj}(c_1 s_{ij}) + c_1 s_{ij} + c_2 \|s_{ij}\|^2 + c_3 s_{ij}^2 \\
    \nonumber
    & \hspace{0.3cm} + \text{conj}(c_3) (\text{conj}(s_{ij}))^2  + c_4 s_{ij}^2 (\text{conj}(s_{ij}))  \\ 
    \nonumber
    & \hspace{0.3cm} + \text{conj}(c_4) s_{ij} (\text{conj}(s_{ij}))^2 + c_5 \|s_{ij}^2\|^2
\end{align}
where $\{c_0,\ldots,c_5\} \subseteq \mathbb{C}$ are coefficients that can be computed from the definition of the Frobenius norm.
Note that the coefficients depend on the pair $(i,j)$ and $M^{[k]}$.
By expressing $s_{ij} = s_{R_{ij}} + \mathrm{i}s_{I_{ij}}$, the objective function $f_1(s_{ij})$ can be written in the following equivalent form:
\begin{align}
    \label{eq:general_fro_2}
    f_2 (\bar{s}_{ij}) =& c_0 + 2 \text{Re}(c_1)s_{R_{ij}} - 2 \text{Im}(c_1)s_{I_{ij}}  \\
    \nonumber
         & + (c_2 + 2 \text{Re}(c_3))s_{R_{ij}}^2 + (c_2 - 2 \text{Re}(c_3))s_{I_{ij}}^2  \\
    \nonumber
         & - 4 \text{Im}(c_3) s_{R_{ij}} s_{I_{ij}} + 2 \text{Re}(c_4) s_{R_{ij}}(s_{R_{ij}}^2 + s_{I_{ij}}^2)  \\
    \nonumber
         & - 2 \text{Im}(c_4) s_{I_{ij}}(s_{R_{ij}}^2 + s_{I_{ij}}^2) + c_5 (s_{R_{ij}}^2 + s_{I_{ij}}^2)^2
\end{align}
where $\bar{s}_{ij} = [s_{R_{ij}}, s_{I_{ij}}]^\text{T}$.
We use the damped newton method (see Algorithm 9.5 in Boyd and Vandenberghe \cite{Boyd2004}) 
to solve the minimization problem.
Therefore, we obtain the local optimum $\bar{s}_{ij}^\star = \text{argmin}_{\bar{s}_{ij} \in \mathbb{R}^2} ~ f_2(\bar{s}_{ij})$ \footnote{ $f_2(\overline{s}_{ij})$ is non-convex for some combinations of the coefficients $\{c_0,\ldots,c_5\}$. Therefore, the solution is only guaranteed to converge to a local optimum.}.
Hence, each $s_{ij}^\star = s_{R_{ij}}^\star + \mathrm{i} s_{I_{ij}}^\star$ has the corresponding scaling matrix $D_{ij}^{\star}$.  
We perform the following update for $k \geq 1$:  \begin{align}
  M^{[k+1]} = D_{ij}^{\star} M^{[k]} (D_{ij}^{\star})^{-1}.
  \label{eq:scaling}
\end{align} 
The iterative algorithm results in the total effective scaling as:
\begin{align}
  D^{''} = \left(\prod_{\forall i, j, i \neq j} D_{ij}^{\star} \right) D_{\mathrm{nr}}^{\star}
  \label{eq:scaling_t}
\end{align} 
where $D_{\mathrm{nr}}^{\star}$ is the optimal diagonal scaling after applying the standard Osborne's iteration.
For example, if we choose to optimize the $s_{12}$ entry then we compute $s_{12}$ by minimizing \eqref{eq:general_fro_2}.
We scale the matrix $M^{[2]} = D_{12}^{\star}M^{[1]}(D_{12}^{\star})^{-1}$, where $M^{[1]} = D_{\mathrm{nr}}^{\star} M (D_{\mathrm{nr}}^{\star})^{-1}$.
We use $M^{[2]}$ and repeat the steps for other $s_{ij}$ until all $s_{ij}$ are computed and effectively $D^{''}$ is obtained.
The above approach allows for computing optimal value of each $s_{ij}$ and then coupling them. 
Finally, the upper bound is computed as $\alpha = \bar{\sigma}(D^{''}M(D^{''})^{-1})$.
We refer to the entire process of computing $D^{''}$ described above as the Generalized Osborne algorithm or GenOsborne for short, which is summarized in Algorithm \ref{alg:GenOsborne}.
\begin{algorithm}[!hbt]
\caption{Upper Bound: GenOsborne Algorithm}
\begin{algorithmic}[1]
  \State (Initialization) Use the standard Osborne's method on $M$ to obtain the  diagonal scaling matrix $D_\mathrm{nr}^\star$. Define $M^{[1]}=D_{\mathrm{nr}}^\star M (D_{\mathrm{nr}}^\star)^{-1}$.
  Set $k=1$.
  %%%%%%%%%%%%%%%%%%%%%%
  \For{$k =1$ to $v(v-1)$} 
  %%%%%%%%%%%%%%%%%%%%%%
  \State Set $(i,j)$
  \State Compute coefficients $\{c_e\}_{e=0}^5$
  for $(i,j)$ and $M^{[k]}$.
  \State Find $\bar{s}_{ij}^\star = \text{argmin}_{\bar{s}_{ij} \in \mathbb{R}^2} ~ f_2(\bar{s}_{ij})$ using the damped newton method and form $s_{ij}^\star = s^\star_{R_{ij}} + \mathrm{i} s^\star_{I_{ij}}$ from $\bar{s}_{ij}^\star$.
  \State Compute the corresponding $D_{ij}^\star$ and set $M^{[k+1]}  = D_{ij}^{\star} M^{[k]} \left( D_{ij}^\star \right)^{-1}$, 
  \EndFor
\State Compute the upper bound $\alpha = \bar{\sigma}(M^{[k]})$
\end{algorithmic}
\label{alg:GenOsborne}
\end{algorithm}
% %%
%%%%%%%%%%%%%%%%%%%%%%%%%%%%%%%%%%%%%%%%%%%
%%%%%%%%%%%%%%%%%%%%%%%%%%%%%%%%%%%%%%%%%%%
%%%
%%%%%%%%%%%%%%%%%%%%%%%%%%%%%%%%%%%%%%%%%%%
%%%%%%%%%%%%%%%%%%%%%%%%%%%%%%%%%%%%%%%%%%%

\end{document}